\documentclass[12pt]{article}
\usepackage{curves}
\usepackage{rotating}
\usepackage{epic}
\usepackage{eepic}
\usepackage{amssymb,amsmath}

\newcommand{\dd}{\mbox{\rm d}}
\newcommand{\DD}{\mbox{\rm D}}

\newcommand{\oo}{\over}
\newcommand{\p}{\partial}

\newcommand{\gam}{\gamma}
\newcommand{\Gam}{\Gamma}

\newcommand{\bV}{\boldsymbol{V}}
\newcommand{\bfs}{\boldsymbol}

\newcommand{\ddg}{\ddagger}

\newcommand{\wg}{\wedge}

\newcommand{\be}{\begin{equation}}
\newcommand{\ee}{\end{equation}}
\newcommand{\bear}{\begin{eqnarray}}
\newcommand{\ear}{\end{eqnarray}}

\newcommand{\lbl}{\label}
\newcommand{\bi}{\bibitem}

\newcommand{\vs}{\vspace}
\newcommand{\hs}{\hspace}
\newcommand{\qb}{\qbezier}
\newcommand{\half}{\mbox{$1\oo 2$}}

\newcommand{\ci}{\cite}

\newcommand{\ellipbz}{
\qbezier(1.0, 0.0)(1.0, 0.4142)(0.7071, 0.7071)
\qbezier(0.7071, 0.7071)(0.4142, 1.0)(0.0, 1.0)
\qbezier(0.0, 1.0)(-0.4142, 1.0)(-0.7071, 0.7071)
\qbezier(-0.7071, 0.7071)(-1.0, 0.4142)(-1.0, 0.0)
\qbezier(-1.0, 0.0)(-1.0, -0.4142)(-0.7071, -0.7071)
\qbezier(-0.7071, -0.7071)(-0.4142, -1.0)(0.0, -1.0)
\qbezier(0.0, -1.0)(0.4142, -1.0)(0.7071, -0.7071)
\qbezier(0.7071, -0.7071)(1.0, -0.4142)(1.0, 0.0)
}

\begin{document}
\baselineskip .59cm

\thispagestyle{empty}

\ 

\vs{1.5cm}

\begin{center}

\baselineskip .95cm

{\bf \LARGE On a Unified Theory of Generalized Branes
Coupled to Gauge Fields, Including the Gravitational and
Kalb-Ramond Fields}

\vs{6mm}

M. Pav\v si\v c

Jo\v zef Stefan Institute, Jamova 39, SI-1000 Ljubljana, Slovenia;
E-mail: matej.pavsic@ijs.si

\vs{1.5cm}

\baselineskip .7cm

ABSTRACT
\end{center}

We investigate a theory in which fundamental objects are branes described
in terms of higher grade coordinates $X^{\mu_1 ... \mu_n}$ encoding
both the motion of a brane as a whole, and its volume evolution.
We thus formulate a dynamics which generalizes the dynamics of the
usual branes. Geometrically, coordinates $X^{\mu_1 ... \mu_n}$ and
associated coordinate frame fields
$\lbrace \gam_{\mu_1 ... \mu_n} \rbrace$ extend the notion of geometry
from spacetime to that of an enlarged space, called Clifford space or
$C$-space. If we start from 4-dimensional spacetime, then the
dimension of $C$-space is 16. The fact that $C$-space has more than
four dimensions suggests that it could serve as a realization of
Kaluza-Klein idea. The ``extra dimensions"
are not just the ordinary extra dimensions, they are related to
the volume degrees of freedom, therefore they are physical, and need
not be compactified. Gauge fields are due to the metric of Clifford space.
It turns out that amongst the latter gauge fields there also exist
higher grade, antisymmetric fields of the Kalb-Ramond type, and their
non-Abelian generalization. All those fields are naturally coupled
to the generalized branes, whose dynamics is given by a generalized
Howe-Tucker action in curved $C$-space.

\vs{1cm}


\section{Introduction}

Point particle is an idealization never found in nature. Physical objects
are extended and possess in principle infinitely many degrees of freedom.
It is now widely accepted that even at the ``fundamental" level objects
are extended. Relativistic strings and higher dimensional extended objects,
branes, have attracted much attention during last three
decades\,\ci{strings,M-theory,BraneWorld,Duff}.

An extended object, such as a brane, during its motion  sweeps 
a worldsheet, whose points form an $n$-dimensional
manifold\footnote{In the literature, 
the name `worldsheet' is often reserved for a 2-dimensional surface swept
by 1-dimensional string. Here we use `worlsheet' for a surface of any dimension
$n$ swept by an $(n-1)$-dimensional brane.
By symbols $V_n$ and $V_N$ we denote {\it manifolds} (we adopt
an old practice), and {\it not} vectors spaces.}
$V_n$ embedded in a target space(time) $V_N$.
Worldsheet is usually considered as being formed by a set of {\it points},
that is, with a worldsheet we associate a {\it manifold of points}, $V_n$.
Alternatively,  we can consider a worldsheet  as being formed by a set of
closed $(n-1)$-branes (that we shall call ``loops"). For instance,
a string world sheet $V_2$ can be considered
as being formed by a set of 1-loops.
In particular such a 1-loop can be just
a closed string which in the course of its evolution sweeps a worldsheet
$V_2$ which, in this case, has the form of a world tube. But in general,
this need not be the case. A set of 1-loops on $V_2$ need not
coincide with a family of strings for various values a time-like
parameter. Thus even a worldsheet swept by an open string can be
considered as a set of closed loops. The ideas that we pursue here are
motivated and based to certain extent by those developed
in refs. \ci{Schild}--\ci{Aurilia}.
We shall
employ the very powerfull geometric language based on Clifford
algebra \ci{Hestenes,Clifford}, which has turned out to be very
suitable for an
elegant formulation of $p$-brane theory and its generalization
\ci{PavsicArena}--\ci{PavsicSaasFee}.
We will employ the property that multivectors of various definite grades,
i.e., $R$-vectors,
since they represent oriented lines, areas, volumes,..., shortly,
$R$-volumes (that we will also call $R$-areas),
can be used in the description of branes. With a
brane one can associate an oriented $R$-volume ($R$-area). Superpositions
of $R$-vectors are generic Clifford
numbers, that we call polyvectors. They represent geometric objects,
which are superposition of oriented lines, areas, volumes,...., that we
associate, respectively, with point particles, closed strings, closed
2-branes,... , or alternatively, with open strings, open 2-branes, open
3-branes, etc.\,. A polyvector is thus used for description of a
physical object, a {\it generalized brane}, whose components are branes
of various dimensionalities.

We thus describe branes by means of higher grade 
coordinates $x^{\mu_1 ...\mu_R}$, $R=0,1,2,...,N$, corresponding to
an oriented $R$-area associated with a brane,
where $N$ is the dimension of the spacetime $V_N$ we
started from. The latter coordinates are {\it collective coordinates},
\ci{AuriliaCastro,AuriliaFuzzy}, analogous to the center of mass coordinates
\ci{PavsicArena}. They do not provide a full description of an extended object,
they merely sample it. Nevertheless, if higher grade coordinates
$x^{\mu_1 ...\mu_R}$ are given, then certainly we have more information
about an extended object than in the case when only its center of
mass coordinates are given. By higher grade coordinates we no longer
approximate an extended object with a point like object; we take into account
its extra structure.

We associate all those higher grade coordinates with {\it points} of
an $2^N$-dimensional space, called {\it Clifford space},
shortly $C$-space, denoted $C_{V_N}$. Every point of $C_{V_N}$ represents
a possible {\it extended event}, associated with a generalized brane.

In order to consider an object's dynamics, one has to introduce a continuous
parameter,
say $\tau$, and consider a mapping $\tau \rightarrow x^{\mu_1...\mu_R} =
X^{\mu_1 ... \mu_R} (\tau)$. So functions $X^{\mu_1 ... \mu_R} (\tau)$
describe a curve in an $2^N$-dimensional space $C_{V_N}$. This generalizes
the concept of {\it worldline} $X^{\mu} (\tau)$ in spacetime $V_N$.
The action principle is given by the {\it minimal length action} in
$C_{V_N}$. That the objetcs, sampled by $X^{\mu_1 ... \mu_R}$ satisfy
such dynamics is our {\it postulate} \ci{PavsicArena,PavsicBook,PavsicSaasFee,
CastroPavsicReview}, we do {\it not} derive it.

The intersection of a $C$-space worldline $X^{\mu_1 ... \mu_R} (\tau)$
with an underlying spacetime $V_N$ (which is a subspace of $C_{V_N}$)
gives, in general an extended event\footnote{Analogously, in spacetime
the intersection
of an ordinary worldline  with a 3-dimensional slice
gives a point.}.
Therefore, what we observe in spacetime are ``instantonic" extended objects
that are localized both in space-like and time-like directions\footnote{
They are the analog of $p=-1$ branes (instantons) that are important
in string theories.}.
According to this generalized dynamics, worldlines are infinitely
extended in $C_{V_N}$, but in general, their
intersections with subspace
$V_N$ are finite.
In spacetime $V_N$ we observe finite 
objects whose time like extension may increase with evolution, and so after
a while they mimic the worldlines of the usual relativity theory.
This has been investigated in refs.\,\ci{PavsicArena,CastroPavsicReview}.
We have also found that such $C$-space theory includes the Stueckelberg
theory \ci{Stueckelberg}--\ci{Pavsic1} as a particular case,
and also has implications for the resolution of the long standing
problem of time in quantum gravity\,\ci{PavsicTime,PavsicBook,PavsicTelAviv}.

Objects described by coordinates $x^{\mu_1 ...\mu_R}$ are {\it points}
in Clifford space $C_{V_N}$, also called {\it extended events}. 
Objects given by functions $X^{\mu_1 ... \mu_R} (\tau)$ are {\it
worldlines} in $C_{V_N}$. A further possibility is to consider, e.g.,
continuous sets of extended events, described by functions
$X^{\mu_1 ... \mu_R} (\xi^A),~A=1,2,...,2^n$, $n < N$, where
$\xi^A \equiv \xi^{a_1...a_r},~r=0,1,2,...,n <N$, are $2^n$ higher grade
coordinates denoting oriented $r$-areas in the parameter space
${\mathbb R}^n$. Functions $X^{\mu_1 ... \mu_R} (\xi^A)$ describe
a $2^n$-dimensional surface in $C_{V_N}$. This generalizes
the concept of {\it worldsheet} or {\it world manifold} $X^\mu (\xi^a),~
a=1,2,...,n$, i.e., the surface that an evolving brane sweeps
in the embedding spacetime $V_N$.

A $C$-space worldline $X^{\mu_1 ... \mu_R} (\tau)$ does not provide a ``full"
description of an extended object, because not ``all" degrees of freedom
are taken into account; $X^{\mu_1 ... \mu_R} (\tau)$ only provides
certain ``collective" degrees of freedom that sample an extended object.
On the contrary, a $C$-space worldsheet $X^{\mu_1 ... \mu_R} (\xi^A)$ provides
much more detailed description, because of the presence of $2^n$ continuous
parameters $\xi^A$, on which the generalized coordinate functions
$X^{\mu_1 ... \mu_R} (\xi^A)$ depend. In particular, the latter
functions can be such that they describe just an ordinary worldsheet,
swept by an ordinary brane. But in general, they describe
more complicated extended objects, with an extra structure.

We equip our manifold $C_{V_N}$ with metric, connection and curvature.
In the case of {\it vanishing curvature}, we can proceed as follows.
We choose in $C_{V_N}$ an {\it origin} ${\cal E}_0$ with coordinates 
$X^{\mu_1 ...\mu_R} ({\cal E}_0) = 0$. This enables us to describe
points ${\cal E}$ of $C_{V_N}$ with {\it vectors} pointing from
${\cal E}_0$ to ${\cal E}$. Since those vectors are Clifford numbers,
we call them {\it polyvectors}. So points of our {\it flat} space
$C_{V_N}$ (i.e., with vanishing curvature) are described by
polyvectors $x^{\mu_1 ...\mu_R} \gam_{\mu_1 ... \mu_R}$, where
$\gam_{\mu_1 ... \mu_R}$ are basis Clifford numbers, that
span a Clifford algebra ${\bfs C}_N$

So our extended objects, the events ${\cal E}$ in $C_{V_N}$, are described
by Clifford numbers. This actually brings {\it spinors} into the
description, since, as is well known, the elements of left (right)
ideals of a Clifford algebra represent spinors\,\ci{IdealSpin}.
So one does not need
to postulate spinorial variables separately, as is usually done in
string and brane theories. Our model is an alternative to the theory of
spinning branes and supersymmetric branes, including spinning strings
and superstrings\,\ci{strings}. In refs.\,\ci{PavsicSaasFee}
it was shown that the
16-dimensional Clifford space provides a framework for a consistent
string theory. One does not need to postulate extra dimensions of
spacetime. One can start from 4-dimensional spacetime, and finds that the
corresponding Clifford space provides enough degrees of freedom
for a string, so that the Virasoro algebra has no central charges.
According to this theory all 16 dimensions of Clifford space are
physical and thus observable \ci{PavsicArena,PavsicKaluza,PavsicKaluzaLong},
because
they are related to the extended nature of objects. Therefore, there is
no need for compactification of the extra dimensions of Clifford space.

As a next step it was proposed\,\ci{PavsicKaluza,PavsicKaluzaLong} that
curved 16-dimensional Clifford space can provide a realization of
Kaluza-Klein theory. Gravitational as well as other gauge interactions
can be unified within such a framework. In ref.\,
\ci{PavsicKaluza,PavsicKaluzaLong} we considered Yang-Mills gauge field
potenitals as components of the $C$-space connection,
and Yang-Mills gauge field strengths as components of the curvature
of that connection.
It was also shown\,\ci{PavsicKaluzaLong} that in a curved $C$-space
which admits $K$ isometries, Yang-Mills gauge potentials occur not only
in the connection, but also in the metric, or equivalently, in the
vielbein. In this paper we concentrate on the latter property, and
further investigate it by studying the brane action in curved background
$C$-space. So we obtain the
minimal coupling terms in the classical generalized brane action,
and we show
that the latter coupling terms contain the ordinary 4-dimensional
gravitational fields, Yang-Mills gauge fields $A_\mu^\alpha$,
$\alpha = 1,2,...,K$, and also the higher grade, in general non-Abelian,
gauge fields $A_{\mu_1 ...\mu_R}^\alpha$ of the Kalb-Ramond type.
We thus formulate an elegant, unified theory for the classical
generalized branes coupled to all those various fields.

\section{On the description of extended objects}

\subsection{Worldsheet described by a set of point events}

As an example of a relativistic extended object, let us first consider
the string. An evolving string  sweeps a worldsheet, a physical
object in the embedding spacetime. Worldsheet can be considered as
being formed by a set of point events. So with a worldsheet we can associate
a 2-dimensional manifold $V_2$, called {\it world manifold},
embedded in an 
$N$-dimensional target space $V_N$. With every {\it point} on $V_2$ we
associate two parameters (coordinates) $\xi^a,~ a = 0,1$ which are arbitrary
(like ``house numbers"). The embedding of $V_2$ into $V_N$ is described
by the mapping
\be
    \xi^a \rightarrow x^\mu = X^\mu (\xi^a) ~, \qquad \xi^a \in R^n \subset
    {\mathbb
R}^n
\lbl{2.1}
\ee
where $x^\mu$, $\mu = 0,1,2,...,N-1$ are coordinates describing position
in $V_N$, whilst $X^\mu (\xi^a)$ are embedding functions (Fig.1),
defined over a domain $R^n$ within a set ${\mathbb R}^n$ of real numbers.
\setlength{\unitlength}{.7mm}
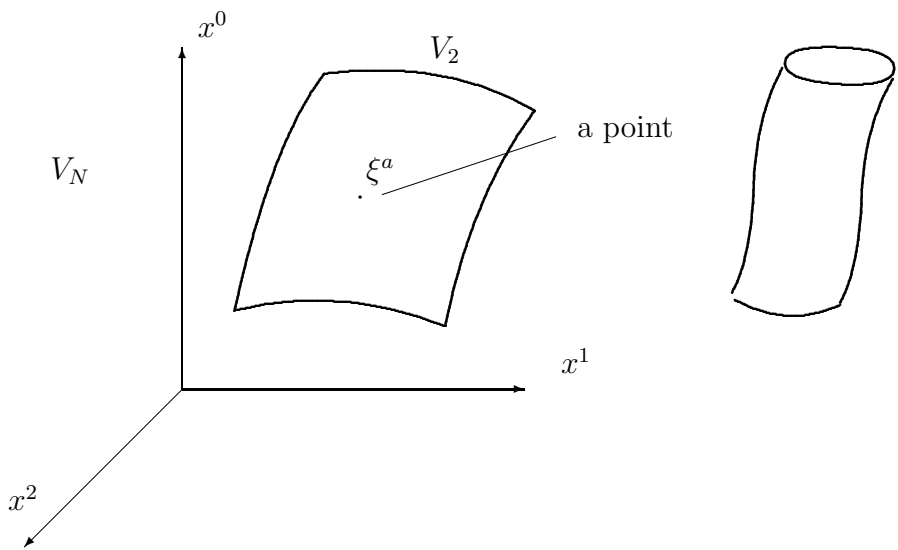
\begin{figure}[h]
\begin{picture}(150,100)(-60,-30)

\put(0,0){\vector(1,0){65}}
\put(0,0){\vector(0,1){65}}
\put(0,0){\vector(-1,-1){30}}

\put(72,3){$x^1$}
\put(-33,-23){$x^2$}
\put(3,67){$x^0$}

\put(35,40){$\xi^a$}
\put(47,63){$V_2$}
\put(33,35){$\cdot$}
\put(-25,40){$V_N$}

\put(38,37){\line(3,1){33}}
\put(75,48){a point}

\thicklines

\qb(10,15)(17,47)(27,60)
\qb(27,60)(50,63)(67,53)
\qb(67,53)(55,37)(50,12)
\qb(50,12)(30,20)(10,15)

\put(125.5,61){\closecurve(9.8,0, 0,4, -11,1, 0,-3)}
\put(155,67){
\renewcommand{\xscale}{10.}
\renewcommand{\yscale}{}
\curve(-5,-50, -4,-53, -3,-51)}

\renewcommand{\xscale}{1.}
\renewcommand{\yscale}{.71}

\put(23.5,0){
\qbezier(81,26)(84,33)(85,50)
\qbezier(85,50)(85,72)(90.5,86) }

\put(44,-2){
\qbezier(81,26)(84,33)(85,50)
\qbezier(85,50)(85,72)(91,86) }

\end{picture}

\caption{A worldsheet can be considered as being formed by a set of
point events associated with {\it points} of a world manifold $V_2$
embedded in $V_N$, described by the
mapping $\xi^a \rightarrow x^\mu = X^\mu (\xi^a)$.}

\end{figure}
 
With our worldsheet we thus associate a 2-parameter set
of {\it points} described by functions $X^\mu (\xi^0,\xi^1)$.
We distinguish here the physical object, the worldsheet, from
the corresponding mathematical object, the manifold (world manifold) $V_2$.
Strictly, we should use two different symbols for those different
objects. In practice, we will not be so rigorous, and we will
simply denote worldsheet by the symbol $V_2$ (in general $V_N$).

\subsection{Worldsheet described by a set of loops}

In previous section a worldsheet was described by a 2-parameter
set of points described by functions $X^\mu (\xi^0,\xi^1)$
Alternatively, instead of points we can consider closed lines, loops,
each being described by functions $X^\mu (s)$, where
$s\in [s_1,s_2] \subset  {\mathbb R}$
is a parameter 
along a loop. A 1-parameter family of such loops $X^\mu (s,\alpha),
~\alpha \in [\alpha_1,\alpha_2] \subset {\mathbb R}$ sweeps a worldsheet
$V_2$. This holds regardless of whether such worldsheet is open or closed.
However, in the case of an open worldsheet, the loops are just
kinematically possible objects, and they cannot be associated with
physical closed strings. In the case of a closed worldsheet, a world tube,
we can consider it as being swept by an evolving closed string.

We will now demonstrate, how with every loop one can associate
an oriented area, whose projections onto the coordinate planes
are $X^{\mu \nu}$. The latter quantities are functionals
of a loop $X^\mu (s)$. If we consider not a single loop, but
a family of loops $X^\mu (s,\alpha),~\alpha \in [\alpha_1,\alpha_2]$,
then $X^{\mu \nu}$ are {\it functions of parameter} $\alpha$, besides being
{\it functionals of a loop}. So we obtain a 1-parameter
family of oriented areas described by functions $X^{\mu \nu} (\alpha)$.
Let us stress again that for every {\it fixed}
$\alpha$, it holds, of course, that $X^{\mu \nu}$ are functionals
of $X^\mu (s,\alpha)$. 

If we choose a loop $B$ on $V_2$, i.e., a loop from a given family
$\lbrace X^\mu (s,\alpha), ~\alpha \in [\alpha_1,\alpha_2] \rbrace$, then we
obtain the corresponding components $X^{\mu \nu}$ of the oriented area
by performing the integration of infinitesimal oriented area
elements over a chosen surface whose boundary is our loop $B$.
Given a boundary loop $B$, it does not matter which surface we choose.
In the following, for simplicity, we will choose just our worldsheet
$V_2$ for the surface.

Let us now consider a surface element
on $V_2$. 
Let $\dd \xi_1 = \dd \xi_1^a \, e_a$ and $\dd \xi_2 = 
\dd \xi_2^a \, e_a$ be
two infinitesimal vectors on $V_2$, expanded in terms of basis
vectors $e_a, ~ a=0,1$. An infinitesimal oriented area is given
by the wedge product
\be
   \dd \Sigma = \dd \xi_1 \wedge \dd \xi_2 = \dd \xi_1^a \, \dd \xi_2^b \,
   e_a \wedge e_b = \mbox{$1\oo 2$} \dd \xi^{ab} \, e_a \wedge e_b
\lbl{2.3}
\ee
where 
\be
    \dd \xi^{ab} = \dd \xi_1^a \, \dd \xi_2^b - 
    \dd \xi_2^a \, \dd \xi_1^b  
\lbl{2.4}
\ee
At every point $\xi \in
V_2$ basis vectors $e_a$ span a 2-dimensional linear vector space,
a tangent space $T_\xi (V_2)$. Following an old tradition (see, e.g.,
\ci{Eisenhart,Rund}) we use symbol $V_n$ for an $n$-dimensional surface embedded
in an $N$-dimensional space $V_N$. Thus $V_n$, and in particular $V_2$,
denotes a manifold, and not a vector space.
In order to simplify our wording, an expression like ``vectors $e_a$ on $V_2$''
will mean ``tangent vectors $e_a$ at a point $\xi \in V_2$". So whenever we
talk about vectors, or whatever geometric objects,
on a manifold (or in a manifold) we just mean that to a given point of the
manifold we attach a geometric object (see, e.g., \ci{Wheeler}). The
latter object, of course, is not an element of our manifold, but of
the tangent space.
Basis vectors on
$V_2$ can be considered as being induced
from the target space basis vectors $\gamma_\mu$:
\be
     e_a = \p_a X^\mu \gamma_\mu
\lbl{2.5}\
\ee

In the following we will adopt the geometric calculus in which basis vectors
are {\it Clifford numbers} satisfying
\be
     \gamma_\mu \cdot \gamma_\nu \equiv \mbox{$1\oo 2$} (\gamma_\mu \gamma_\nu
     + \gamma_\nu \gamma_\mu) = g_{\mu \nu}
\lbl{2.6}
\ee
where $g_{\mu \nu}$ is the metric of $V_N$. Eq. (\ref{2.6}) defines
the {\it inner
product} of two vectors as the symmetric part of the Clifford product
$\gamma_\mu \gamma_\nu$. The antisymmetric part of $\gamma_\mu \gamma_\nu$
is identified with the {\it wedge} or {\it outer product}
\be
     \gamma_\mu \wedge \gamma_\nu \equiv \mbox{$1\oo 2$} (\gamma_\mu \gamma_\nu
     - \gamma_\nu \gamma_\mu)
\lbl{2.6a}
\ee 
Analogous relations we have for the worldsheet basis vectors $e_a$:
\be
      e_a \cdot e_b \equiv \mbox{$1\oo 2$}(e_a e_b + e_b e_a) = \gam_{ab}
\lbl{2.6b}
\ee
\be
     e_a \wg e_b
\equiv \mbox{$1\oo 2$}(e_a e_b - e_b e_a)
\lbl{3.6c}
\ee
where $\gam_{ab}$ is the metric on $V_2$ which, according to eq.\,(\ref{2.5}),
can be considered as being induced from the target space.

If we insert the relation (\ref{2.5}) into eq.(\ref{2.3}) we have
\be
    \dd \Sigma = \mbox{$1\oo 2$} \dd \xi^{ab} \, \p_aX^\mu \p_b X^\nu
    \, \gamma_\mu \wedge \gamma_\nu
\lbl{2.7}
\ee
This is an infinitesimal {\it bivector} or {\it 2-vector} in the target
space $V_N$.

A finite 2-vector is obtained upon integration\footnote{Such integration
poses no problem in {\it flat} $V_N$. In curved space we may still use
the same expression (\ref{2.8}) which then {\it defines} such integral
that all vectors are carried together by means of parallel transport along
geodesics into a chosen point of $V_N$ where the integration is
actually performed \ci{PavsicOrder}.} over a finite surface $\Sigma_B$
enclosed by
a loop $B$:
\bear
    \int_{\Sigma_B}
     \dd \Sigma \equiv \half X^{\mu \nu} \, \gamma_\mu \wedge \gamma_\nu
    &=& {1\oo 2} \int_{\Sigma_B} \dd \xi^{ab} \p_aX^\mu \p_b X^\nu \, 
    \gamma_\mu \wedge \gamma_\nu \nonumber \\
    &=&{1\oo 2} \int_{\Sigma_B} \dd \xi^{ab} {1\oo 2} 
    (\p_aX^\mu \p_b X^\nu -
    \p_aX^\nu \p_b X^\mu) \gamma_\mu \wedge \gamma_\nu
\lbl{2.8}
\ear
From eq.(\ref{2.8}) we read
\be
     X^{\mu \nu} [B] =  {1\oo 2} \int_{\Sigma_B} \dd \xi^{ab}
     (\p_aX^\mu \p_b X^\nu -
    \p_aX^\nu \p_b X^\mu) 
\lbl{2.9}
\ee
By Stokes theorem this is equal to
\be
      X^{\mu \nu} [B] = {1\oo 2} \oint_B \dd s \left ( X^\mu \, 
      {{\p X^\nu}\oo 
      {\p s}} -  X^\nu \, {{\p X^\mu}\oo {\p s}} \right )
\lbl{2.10}
\ee
where $X^\mu (s)$ describes a boundary loop $B$, $s$ being a parameter along
the loop.

Eq.(\ref{2.10})
demonstrates that $X^{\mu \nu}$ are components of the bivector,
determining an oriented area, associated with a surface enclosed
by a loop $X^\mu (s)$ on the worldsheet $V_2$. Hence there is a close
correspondence between surfaces and the boundary loops.
The components  $X^{\mu \nu}$ can be therefore be considered as bivector
coordinates
of a loop. These are collective coordinates, since
the detailed shape (configuration) of the loop
is not determined by $X^{\mu \nu}$. Only the oriented area associated with
a surface enclosed by the loop is determined by $X^{\mu \nu}$.
Therefore $X^{\mu \nu}$ refers to a class of loops, from which we
may choose a {\it representative loop}, and say that $X^{\mu \nu}$ are its
coordinates.
From now on, `loop' we will be often a short hand expression
for a {\it representative loop} in the sense above.

By means of eqs.\,(\ref{2.8})--(\ref{2.10}) we have performed a mapping from
an {\it infinite dimensional space} of loops $X^\mu (s)$ into a {\it finite
dimensional space} of oriented areas $X^{\mu \nu}$. Instead of
describing loops by infinite dimensional objects $X^\mu (s)$, we can
describe them by finite dimensional objects, oriented areas, with
bivector coordinates $X^{\mu \nu}$. {\it We have thus arrived at
a finite dimensional description of loops} (in particular, closed strings),
the so called {\it quenched minisuperspace description} suggested
by Aurilia et al.\,\ci{AuriliaCastro}.

When we consider not a single loop $X^\mu (s)$, but a 1-parameter
family of loops $X^\mu (s,\alpha)$, we have a worldsheet, considered
as being formed by a set of loops. By means of eqs.\,(\ref{2.8})--(\ref{2.10}),
with every loop within such a family, i.e., for a fixed $\alpha$, we can
associate bivector coordinates $X^{\mu \nu}$. For variable $\alpha$
we then obtain functions $X^{\mu \nu} (\alpha)$. This is a quenched
minisuperspace description of a worldsheet. A full description is
in terms of embedding functions $X^\mu (\xi^0,\xi^1)$, or a family
of loops $X^\mu (s,\alpha)$.

In the following we will consider two particular choices for parameter
$\alpha$.

In eq.(\ref{2.8}) we have the expression for an oriented area
associated with a loop $B$. It has been
obtained upon the integration of the infinitesimal oriented surface
elements (\ref{2.3}). 
Besides the oriented area we can associate
with our loop on $V_2$ also a {\it scalar}
quantity, namely the {\it scalar area} ${\cal A}$ which we obtain
according to
\be
     {\cal A} = \int_{\Sigma_B} \sqrt{\dd \Sigma^{\ddg} \cdot \dd \Sigma}
\lbl{2a.1}
\ee
Here `$\ddg$' denotes
{\it reversion}, that is the operation which reverse the order of vectors
in a product.
Using the relation $e_a \wedge e_b = e \epsilon_{ab}$, where
 $e = e_1 \wedge 
e_2$ is the pseudoscalar in 2-dimensional space $V_2$ such that
$e^{\ddg} \cdot e = (e_2 \wedge e_1) \cdot (e_1 \wedge e_2) =
\gamma_{11} \gamma_{22} - \gamma_{21} \gamma_{12} = {\rm det} \, \gamma_{ab}
\equiv \gamma$, we find  
\be
     \dd \Sigma^{\ddg} \cdot \dd \Sigma = {1\oo 4} \gamma (\dd \xi^{ab}
     \epsilon _{ab})^2
\lbl{21.2}
\ee 
i.e.
\be
    {\cal A} = \int \sqrt{\dd \Sigma^{\ddg} * \dd \Sigma} =
    {1\oo 2}  \int \sqrt{|\gamma|} \dd \xi^{ab} \epsilon_{ab}
    = \int \sqrt{|\gamma|} \, \dd \xi^{12}
\lbl{2a.3}
\ee
If we choose $\dd \xi_1^a = (\dd \xi^1,0), ~ \dd \xi_2^a = (0, \dd \xi^2)$,
then $\dd \xi^{12} = \dd \xi^1 \dd \xi^2$.

We imagine that our surface $V_2$ is covered by a family of
loops $X^\mu (\alpha,s)$
(Fig.2), such that the totality of points of all those loops is in
one-to-one correspondence with the points of the manifold $V_2$,
and that for parameter $\alpha$ we take the scalar area ${\cal A}$.
To every loop there belong bivector coordinates $X^{\mu \nu}$
(calculated according to eq.(\ref{2.8})) and a scalar parameter ${\cal A}$
(calculated according to eq. (\ref{2a.3})), determining the the scalar area.
Dependence of $X^{\mu \nu}$ on ${\cal A}$ is characteristic for a given
class of surfaces $V_2$. If we consider a different class of surface
$V_2$, then functions $X^{\mu \nu} ({\cal A})$ are in general different
(Fig. 3)

\setlength{\unitlength}{.7mm}
\begin{figure}[h]
\begin{picture}(150,100)(-60,-30)

\put(0,0){\vector(1,0){65}}
\put(0,0){\vector(0,1){65}}
\put(0,0){\vector(-1,-1){30}}

\put(72,3){$x^1$}
\put(-33,-23){$x^2$}
\put(3,67){$x^0$}

\put(35,38){${\cal A}$}
\put(47,65){$V_2$}
\put(-25,40){$V_N$}
\put(112,22){${\cal A}$}

\put(44,37){\line(3,1){28}}
\put(75,46){a 1-loop}
\put(110,35){\line(-1,1){12}}


\thicklines

\begin{rotate}{-10}
\renewcommand{\xscale}{.38}
\renewcommand{\yscale}{.45}
\put(30,44){\closecurve(0,-24, 20,0, 0,19, -22,0)}
\end{rotate}

\renewcommand{\xscale}{1.}
\renewcommand{\yscale}{1.}

\qb(10,15)(17,47)(27,60)
\qb(27,60)(50,63)(67,53)
\qb(67,53)(55,37)(50,12)
\qb(50,12)(30,20)(10,15)

\renewcommand{\xscale}{2.2}
\renewcommand{\yscale}{1.3}
\put(126.,59.){\closecurve(9.8,0, 0,4, -11,1, 0,-3)}
\put(145,87){
\renewcommand{\xscale}{8.}
\renewcommand{\yscale}{1.5}
\curve(-4.6,-50, -4,-53, -3,-50)}

\put(191,77){
\renewcommand{\xscale}{19.}
\renewcommand{\yscale}{.8}
\curve(-4.6,-50, -4,-55, -3,-50)}

\begin{rotate}{14}
\renewcommand{\xscale}{1.}
\renewcommand{\yscale}{1.3}

\put(22.5,-79){
\qbezier(85,50)(85,72)(90.5,86) }
\end{rotate}

\begin{rotate}{-16}
\renewcommand{\xscale}{1.}
\renewcommand{\yscale}{.81}
\put(32,25){
\qbezier(80,25)(84,33)(85,50)
\qbezier(85,50)(85,72)(91,86) }
\end{rotate}

\end{picture}

\caption{We consider a loop on $V_2$. It determines an {\it oriented
area} whose extrinsic 2-vector coordinates are $X^{\mu \nu}$. The {\it scalar
area} of the surface element enclosed by the loop is ${\cal A}$. Given an
initial loop, functions $X^{\mu \nu} ({\cal A})$ are characteristic
for a class of the surface $V_2$.}

\end{figure}
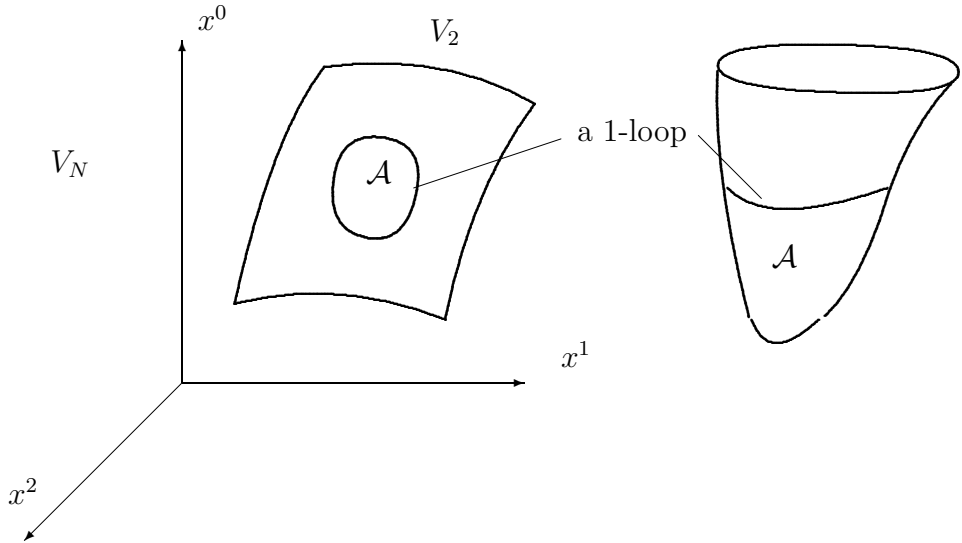

\setlength{\unitlength}{.6mm}

\begin{figure}[h]
\hs{3mm} \begin{picture}(100,40)(-50,0)

\put(10,15){
\renewcommand{\xscale}{7.}
\renewcommand{\yscale}{2.2}
 \qbezier(-1.0, 0.0)(-1.0, -0.4142)(-0.7071, -0.7071)
\qbezier(-0.7071, -0.7071)(-0.4142, -1.0)(0.0, -1.0)
\qbezier(0.0, -1.0)(0.4142, -1.0)(0.7071, -0.7071)
\qbezier(0.7071, -0.7071)(1.0, -0.4142)(1.0, 0.0)}

\put(10,40){
\renewcommand{\xscale}{7.}
\renewcommand{\yscale}{2.2}
\ellipbz }

\put(3,15){\line(0,1){25}}

\put(17,15){\line(0,1){25}}

\put(30,15){
\renewcommand{\xscale}{7.}
\renewcommand{\yscale}{2.2}
 \qbezier(-1.0, 0.0)(-1.0, -0.4142)(-0.7071, -0.7071)
\qbezier(-0.7071, -0.7071)(-0.4142, -1.0)(0.0, -1.0)
\qbezier(0.0, -1.0)(0.4142, -1.0)(0.7071, -0.7071)
\qbezier(0.7071, -0.7071)(1.0, -0.4142)(1.0, 0.0) }

\put(38.5,32){
\renewcommand{\xscale}{7.}
\renewcommand{\yscale}{2.2}
\ellipbz }

\put(23,15){\line(1,2){9}}
\put(37,15){\line(1,2){8.5}}

\put(90,40){
\renewcommand{\xscale}{12.7}
\renewcommand{\yscale}{4.}
\ellipbz }

\put(90,15){\line(1,2){12}}
\put(90,15){\line(-1,2){12}}

\put(128,32){
\renewcommand{\xscale}{12.5}
\renewcommand{\yscale}{4.}
\ellipbz }

\put(110,15){\line(92,47){30}}
\put(110,15){\line(3,10){5.4}}

\put(15,0){(a)}
\put(100,0){(b)}

\end{picture}

\caption{ (a) Examples of two different surfaces belonging to a class
of surfaces that all satisfy
equation $X^{12} ({\cal A}) = k$. Constant $k$
differs from zero, if a cillindric surface has only one boundary loop,
so that, e.g., the upper part is open, whilst the lower part is closed.
(b) Example of surfaces belonging to a class of surfaces that satisfy equation
$X^{12} ({\cal A}) = k {\cal A}$. }

\end{figure}
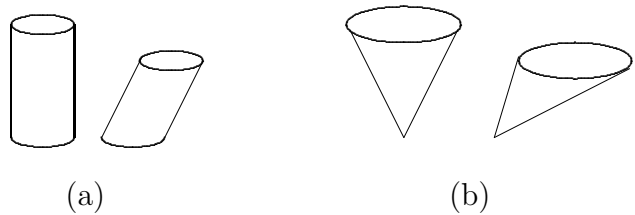

Instead of starting with a given surface $V_2$ on which we determine a family
of loops and calculate the functions $X^{\mu \nu} ({\cal A})$, we can start
from the other end. We may assume that all what is known are functions 
$X^{\mu \nu} ({\cal A})$. From those functions we do not know what
the surface (worldsheet)  $V_2$ exactly
is, but we have some partial information (see Fig.\,3), up to a class
of surfaces (worldsheets). Functions
$X^{\mu \nu} ({\cal A})$ provide a means of describing  a surface $V_2$,
although not in all details.

Instead of the scalar area ${\cal A}$ we may take as the parameter
any other parameter\footnote{In a suitable 
choice of parameters $\xi^a$ the determinant of the metric $\gamma_{ab}$
on $V_2$ can be constant, that is $\p_a |\gamma| = 0$. Choosing
$|\gamma| = 1$, the scalar area ${\cal A}$ is given just by the integral
(\ref{2.11}).}.
We may take, for instance, just the integral\footnote{Similarly,
in describing a world line
$X^\mu (\tau)$, we may take any parameter $\tau$ which, in particular,
can be the length of the worldline. The analog of eq.\,(\ref{2.11}) is
$\tau = \int_0^{\tau} \dd \tau'$.}
in the parameter space ${\mathbb R}^2$
\be
     \xi^{ab} (B) = \int_{\Sigma_B} \dd \xi'^{ab}
\lbl{2.11}
\ee
taken over a domain corresponding to a surface $\Sigma_B$ with boundary $B$.

So we have a mapping $B \rightarrow 
\xi^{ab} (B)$, such that to any boundary loop $B$ of our family
there correspond parameters $\xi^{ab}$.
Because of the property
$\xi^{ab} = - \xi^{ba}$, (where $a,b = 0,1$ if $V_2$ is time like),
there is in fact
a single parameter $\xi^{12}$.
The extrinsic 2-vector coordinates $X^{\mu \nu}$ are functions of 
$\xi^{ab}$. The mapping
\be
\xi^{ab} \rightarrow x^{\mu \nu} = X^{\mu \nu} (\xi^{ab})
\lbl{2.11aa}
\ee
determines a class of surfaces $V_2$, embedded in $V_N$, which are
all in accordance with eq.\,(\ref{2.11aa}). Knowing the functions
$X^{\mu \nu} (\xi^{ab})$ means knowing a class $\lbrace V_2 \rbrace$,
but not a particular $V_2 \in \lbrace V_2 \rbrace$.

The mapping $\tau \rightarrow x^\mu = X^\mu (\tau)$,
involving vector coordinates, describes a curve (a ``worldline")
in the space spanned by vectors $\gam_\mu$.
The derivative of $X^\mu$ with respect to $\tau$, i.e., ${\dot X}^\mu =
\dd X^\mu/\dd \tau$, is the tangent vector to the worldline, or 
{\it velocity}.

Similarly, the mapping (\ref{2.11aa}), involving bivector coordinates,
describes a curve (a ``worldline'') in the space spanned by the bivectors
$\gam_\mu \wg \gam_\nu$, and we can calculate the derivative
\be
    \p_{ab} X^{\mu \nu} \equiv \frac{\p X^{\mu \nu}}{\p \xi^{ab}}
\lbl{2.12}
\ee
which generalizes the concept of velocity.

\subsection{Generalization to arbitrary dimensions}

Let us now consider extended objects associated with manifolds
$V_n$ that have arbitrary dimension $n$ and are embedded in a target space
$V_N$ of dimension $N$. An infinitesimal infinitesimal oriented
area element on $V_n$ is an
$n$-vector
\be
      \dd \Sigma = \dd \xi_1 \wedge \dd \xi_2 \wedge ... 
      \wedge \dd \xi_n =
      \dd \xi_1^{a_1} \dd \xi_2^{a_2} ... \dd \xi_n^{a_n} e_{a_1} \wedge
      e_{a_2} \wedge ... \wedge e_{a_n} = {1\oo n!} \dd \xi^{a_1 ... a_n}
      e_{a_1} \wedge ... \wedge e_{a_n}
\lbl{2.23}
\ee
where
\be
     \dd \xi^{a_1 ... a_n} = \dd \xi_1^{[a_1} \dd \xi_2^{a_2} ... 
     \dd \xi_n^{a_n]}
\lbl{2.24}
\ee
If we consider the basis vectors $e_a$ on $V_n$ as being induced
from the basis
vectors $\gamma_\mu$ of the embedding space $V_N$ according to the relation
(\ref{2.5}), we have
\be
      \dd \Sigma = {1\oo {n!}} \dd X^{\mu_1 ... \mu_n} \gamma_{\mu_1}
      \wedge ... \wedge \gamma_{\mu_n} =
      {1\oo {n!}} \dd \xi^{a_1 ... a_n} \p_{a_1} X^{\mu_1} ... \p_{a_n}
      X^{\mu_n}  \gamma_{\mu_1} \wedge ... \wedge \gamma_{\mu_n}  
\lbl{2.25}
\ee
After the integration over a finite $n$-surface $\Sigma_B$
with boundary $B$
we obtain a finite $n$-vector
\bear
     \int_{\Sigma_B} \dd \Sigma &=&  {1\oo {n!}} X^{\mu_1 ... \mu_n} 
     \gamma_{\mu_1}
      \wedge ... \wedge \gamma_{\mu_n} =  
      {1\oo {n!}} \int_{\Sigma_B} \dd \xi^{a_1 ... a_n}
      \p_{a_1} X^{\mu_1} 
      ... \p_{a_n}
      X^{\mu_n}  \gamma_{\mu_1} \wedge ... \wedge \gamma_{\mu_n} 
      \nonumber \\ 
      &=& {1\oo {n!}} \int_{\Sigma_B} \dd \xi^{a_1 ... a_n} {1\oo {n!}}
      \p_{[a_1} X^{\mu_1} ... \p_{a_n]}
      X^{\mu_n}  \gamma_{\mu_1} \wedge ... \wedge \gamma_{\mu_n}  
\lbl{2.26}
\ear
Its $n$-vector components are
\be
     X^{\mu_1 ... \mu_n} [B] =\int_{\Sigma_B} \dd \xi^{a_1 ... a_n} 
     {1\oo {n!}} \p_{[a_1} X^{\mu_1} ... \p_{a_n]} X^{\mu_n}
\lbl{2.27}
\ee
They describe an oriented $n$-area associated with $\Sigma_B$, whose
boundary $B$ will be called $(n-1)$-loop,
and  $X^{\mu_1 ... \mu_n} [B]$ are its {\it extrinsic coordinates}.
With the same $(n-1)$-loop we can associate {\it intrinsic coordinates}
(parameters), in analogy to
eqs.\,(\ref{2a.1})--(\ref{2.11}), according to
\be
     \xi^{a_1 ... a_n} (B) = \int_{\Sigma_B} \dd \xi'^{a_1 ... a_n}
\lbl{2.28}
\ee
With a particular choice of coordinates $\xi^a$, such that
det $\gam_{ab} = 1$, the quantities $\xi^{a_1 ...a_n}$ in the above
equation determine the intrinsic (scalar) $n$-area of the $n$-surface
bounded by the $(n-1)$-loop.

As in the case of $V_2$ we assume that on our $n$-dimensional
worldsheet $V_n$ there exists a family of $(n-1)$-loops $B$, described
by functions $X^\mu (s^{\bar a},\alpha),~{\bar a} = 1,2,...,n-1$,

Instead of a manifold $V_n$ of points we thus consider a
family of loops.
With every $(n-1)$-loop of the family we can associate arbitrary  parameters
$\xi^{a_1 ... a_n}$
(coordinates are like ``house numbers"). Because of the property
\be
     \xi^{a_1 ... a_j a_k ... a_n} = - \xi^{a_1 ... a_k a_j ... a_n}
\lbl{2.30a}
\ee
there is actually a single parameter.
This is a particular choice for
parameter $\alpha$ of our family of loops $X^\mu (s^{\bar a},\alpha)$.
By means of a mapping
\be
    \xi^{a_1 ... a_n} \rightarrow x^{\mu_1 ... \mu_n} =
    X^{\mu_1 ... \mu_n} (\xi^{a_1 ... a_n})
\lbl{2.30}
\ee
we obtain a quenched minisuperspace description of a family of $(n-1)$-loops,
i.e., a description in terms of the target space
multivector coordinate functions $X^{\mu_1 ... \mu_n}$.
The family is such
that the totality of the points of the $(n-1)$-loops belonging to the
family is in one-to-one correspondence with the points of
the worldsheet $V_n$.
In other words, by mapping (\ref{2.30}) we have {\it a quenched minisuperspace
description of worldsheet}. 

We started from a brane described by the embedding functions
$X^\mu (\xi^a )$,
and derived the expression (\ref{2.27}) and functions (\ref{2.30}).
Once we have
$X^{\mu_1 ... \mu_2}$ as functions of a parameter $\xi^{a_1 ... a_n}$, we
may {\it forget} about the embedding $x^\mu = X^{\mu} (\xi^a)$ that we
started from. {\it We may assume that all the information available to us
are just functions $X^{\mu_1 ...\mu_n} (\xi^{a_1 ... a_n})$
given by mapping (\ref{2.30})}. Then we do not have knowledge of
a particular worldshet's
manifold
$V_n$, but of a class $\lbrace V_n \rbrace$ of worldsheet's manifolds that all
satisfy eq.\, (\ref{2.30}) for given functions
$X^{\mu_1 ... \mu_n} (\xi^{a_1 ... a_n})$.
So we calculate the derivative
\be
    \p_{a_1 ... a_n} X^{\mu_1 ... \mu_n} \equiv
    {{\p X^{\mu_1 ... \mu_n}}\oo {\p \xi^{a_1 ... a_n}}}
\lbl{2.29}
\ee
which generalizes the notion of velocity.

In the above considerations one has to bear in mind that many loop
configuration may cast the same holographic projections onto the
coordinate planes as a single loop configuration. Therefore, a given
set of polyvector coordinates $X^{\mu_1 ...\mu_n} (\xi^{a_1 ... a_n})$
may describe a single or many loop configuration. Not only the details
of a single loop configuration (its infinite dimensionality), but also
the number of loops is undetermined in this quenched description
of loops.

\section{The dynamics of extended objects}

\subsection{Objects described in terms of $X^\mu (\xi^a)$}

The extended objects described by the mapping (\ref{2.1}) obey the dynamical
law that is incorporated in the Dirac-Nambu-Goto minimal surface action.
An equivalent action that was considered in ref. \ci{PavsicBook} is a functional
of the embedding functions $X^\mu (\xi^a)$ and the coordinate basis vector
fields
$e^a (\xi)$ having the role of Lagrange multipliers:
\be
     I[X^\mu, e^a] = {\kappa \oo 2} \int \dd^n \xi \, |e| \, (e^a 
     \p_a X^\mu \, e^b \p_b X_\mu + 2 - n)
\lbl{2.31}
\ee
where $|e| \equiv \sqrt{|\gamma|}$ is the determinant of $\gamma_{ab}
= e_a \cdot e_b$.

Expanding the coordinate vector fields $e^a (\xi)$, $a=1,2,...,n$ in terms
of orthonormal vector fields\footnote{Their inner products 
$e^{\bfs  a} \cdot e^{\bfs  b}
= \eta^{{\bfs  a}{\bfs  b}}$ gives the Minkowski metric.}
$e^{\bfs  a}$, ${\bfs  a} = 1,2,...,n$, by means of a tetrad
${e^a}_{\bfs  a} (\xi)$ according to
\be
    e^a (\xi) = {e^a}_{\bfs  a} (\xi) e^{\bfs  a}
\lbl{2.32}
\ee
we find the following relations\footnote{Eq.(\ref{2.33}) also comes directly
from the relation for a derivative with respect to a vector \ci{Hestenes}.}
\be
     {{\p e^a}\oo {\p e^b}} \equiv e_{\bfs  c} \, 
     \frac{\p e^a}{\p {e^b}_{\bfs  c}}
     = n {\delta^a}_b
\lbl{2.33}
\ee
\be
     {{\p |e|}\oo {\p e^a}} = {{\p |e|}\oo {\p \gamma^{cd}}} \,
     {{\p \gamma^{cd}}\oo {\p e^a}} = - n \, |e| e_a
\lbl{2.34}
\ee
where $n$ comes from the contraction 
$e^{\bfs  a} e_{\bfs  a} = n$.

Using (\ref{2.33}),(\ref{2.34}) we find that the variation of the action
(\ref{2.31}) with respect to $e^a$ gives
\be
    - {1\oo 2} e_c (e^a \p_a X^\mu \, \p^b X_\mu + 2 - n) +
    \p_c X^\mu \, \p_d         X_\mu e^d = 0
\lbl{2.35}
\ee
Performing the inner product with $e^c$ and using $e^c \cdot e_c = n$ we
find
\be
    e^a \p_a X^\mu \, e^b \p_b X_\mu = n
\lbl{2.36}
\ee
and eq.(\ref{2.35}) becomes
\be
   e_c = \p_c X^\mu \, \p_d X_\mu e^d
\lbl{2.37}
\ee
This is the equation of ``motion" for the Lagrange multipliers $e_a$. 
In order
to understand better the meaning of eq.(\ref{2.37}) let us perform the
inner product with $e_a$:
\be
     e_c \cdot e_a = \p_c X^\mu \, \p_d X_\mu e^d   \cdot e_a
\lbl{2.38}
\ee
Since $e_c \cdot e_a = \gamma_{ca}$ and $e^d \cdot e_a = {\delta^d}_a$ we
obtain after renaming the indices
\be
     \gamma_{ab} = \p_a X^\mu \, \p_b X_\mu
\lbl{2.39}
\ee
This is the relation for the induced metric on the worldsheet.
On the other hand, eq.\,(\ref{2.38}) can be written as
\be
     e_a \cdot e_b = (\p_a X^\mu \gam_\mu) \cdot (\p_a X^\nu \gam_\nu)
\lbl{2.38a}
\ee
from which we have that
basis vectors $e_a$ on the worldsheet $V_n$
are expressed in terms of the embedding space basis vectors $\gam_\mu$:
\be
    e_a = \p_a X^\mu \gam_\mu
\lbl{2.38b}
\ee
With our procedure we have thus derived eq.\,(\ref{2.5}) as a solution
to our dynamical sytem.

Using eq.\,(\ref{2.39}) we find that
the action (\ref{2.31}) is equivalent to the well known Howe--Tucker
action which is a functional of $X^\mu (\xi)$ and $\gamma^{ab}$:
\be
    I[X^\mu,\gam_{ab}] = \frac{\kappa}{2} \int \dd^n \xi \,
    \sqrt{|\gam|} \, (\gam^{ab} \p_a X^\mu \p_b X_\mu + 2-n)
\lbl{2.39a}
\ee

\subsection{Objects described in terms of
$X^{\mu_1 ... \mu_n} (\xi^{a_1 ... a_n})$}

In Sec.\,2.3
we have seen that an alternative description of extended objects, up
to a class in which all objects have the same coordinates
$X^{\mu_1 ...\mu_n}$, is given by
the mapping (\ref{2.30}). Let us assume that such objects are described
by the following action
   $$I[X^{\mu_1, ..., \mu_n},e] = {\kappa\oo 2} \int \dd^n \xi \, |e| \,
   \Biggl[ {1 \oo {n!}} \left ( {1\oo {n!}} \, e^{a_1} \wedge ... \wedge
   e^{a_n} \, {{\p X^{\mu_1 ... \mu_n}}\oo {\p \xi^{a_1 ... a_n}}}
   \right )^{\ddagger} \hs{3cm}$$
\be   
    \hs{4cm} \times \left ({1\oo {n!}} \, e^{b_1} \wedge ... \wedge
   e^{b_n} \, {{\p X_{\mu_1 ... \mu_n}}\oo {\p \xi^{b_1 ... b_n}}}
   \right ) + 1 \Biggr]
\lbl{2.35a}
\ee
Factor $1/n!$ inside the bracket comes from the definition of the worldsheet
$n$-vector $(1/n!)e^{a_1} \wedge ... \wedge e^{a_n} \p X^{\mu_1 ... \mu_n}/
\p \xi^{a_1 ... a_n}$. The extra factor $1/n!$ in front of the bracket
comes from the square of the target space $n$-vector
$(1/n!) (\p X^{\mu_1 ... \mu_n}/\p \xi^{a_1 ... a_n} )
\gamma_{\mu_1} \wedge
... \wedge \gamma_{\mu_n}$. The operation $\ddagger$ reverses the order
of vectors.

Let us take into account the following relations:
\be
    e_{a_1} \wedge ... \wedge e_{a_n} = e \, \epsilon_{a_1 ... a_n}
\lbl{2.36a}
\ee
\be
    e^{a_1} \wedge ... \wedge e^{a_n} = e^{-1} \, \epsilon^{a_1 ... a_n}
\lbl{2.37a}
\ee
\be
    e^{-1} = {e \oo {|e|^2}} \; , 
    \qquad |e|\equiv \sqrt{e^{\ddg} \cdot e}
    = \sqrt{|\gamma|} \equiv \lambda
\lbl{2.38aa}
\ee
\be
     \gamma = {\rm det} \, \gamma_{ab} \; , \qquad \gamma_{ab} = 
     e_a \cdot e_b
\lbl{2.40a}
\ee
Instead of the intrinsic parameters $\xi^{a_1 ...a_n}$, let 
us introduce the dual parameter
\be
    {\tilde \xi} = {1\oo {n!}} \, \epsilon_{a_1 ... a_n} \xi^{a_1 ... a_n}
\lbl{2.40}
\ee
and rewrite the $n$-area velocity according to
\be
    {{\p X^{\mu_1 ... \mu_n}}\oo {\p \xi^{c_1 ... c_n}}} =
    {{\p X^{\mu_1 ... \mu_n}}\oo {\p {\tilde \xi}}} {{\p {\tilde \xi}}\oo
    {\p \xi^{c_1 ... c_n}}} = {{\p X^{\mu_1 ... \mu_n}}\oo {\p {\tilde \xi}}}
    \epsilon_{c_1 ... c_n}
\lbl{2.41}
\ee
where we have used
\be
      {{\p \xi^{a_1 ... a_n}}\oo {\p \xi^{c1 ... c_n}}} =
      \delta_{c_1 ... c_n}^{a_1 ... a_n}
\lbl{2.42}
\ee
and where the generalized Kronecker symbol is given by the antisymmetrized
sum of products of ordinary deltas.
From eq.(\ref{2.41}) we have
\be
     {{\p X^{\mu_1 ... \mu_n}}\oo {\p {\tilde \xi}}} = {1\oo {n!}} \,
     \epsilon^{c_1 ... c_n} \,  {{\p X^{\mu_1 ... \mu_n}}\oo 
     {\p \xi^{c_1 ... c_n}}}
\lbl{2.43}
\ee
By using eqs.((\ref{2.36}),(\ref{2.38}) and (\ref{2.43}) we can rewrite
the action (\ref{2.35a}) as
\be
    I[X^{\mu_1 ...\mu_n},\lambda] = {\kappa\oo 2} \int \dd {\tilde \xi} \,
    \left ( {1\oo {\lambda \, n!}} 
    {{\p X^{\mu_1 ...\mu_n}}\oo {\p {\tilde \xi}}}
    {{\p X_{\mu_1 ...\mu_n}}\oo {\p {\tilde \xi}}} + \lambda \right )
\lbl{2.44}
\ee
where
\be
     \dd {\tilde \xi} = {1\oo n!} \epsilon_{a_1 ... a_n} \dd 
     \xi^{a_1 ... a_n}
     = {1\oo n!} \epsilon_{a_1 ... a_n} \, \dd \xi_a^{[a_1} ... \dd 
     \xi_n^{a_n]} = \epsilon_{a_1 ... a_n} \, \dd \xi_a^{a_1} ... \dd 
     \xi_n^{a_n} = \dd \xi^1 \dd \xi^2 ... \dd \xi^n \equiv \dd^n \xi
\lbl{2.45}
\ee
The last step in eq.(\ref{2.45}) holds in a coordinates system in which
$\dd \xi_1^{a_1} = \dd \xi^1$, $\dd \xi_2^{a_2} = \dd \xi^2$, ..., $\dd 
\xi_n^{a_n} = \dd \xi^n$.

The action (\ref{2.44}) is a functional of the $n$-area variables
$X^{\mu_1 ... \mu_n} ({\tilde \xi})$ and a Lagarange multiplier 
$\lambda$, defined
in eq.(\ref{2.38aa}). Variation of eq. (\ref{2.44}) with respect to $\lambda$
and $X^{\mu_1 ... \mu_n}$, respectively, gives
\bear
     \delta \lambda &:& \quad {1\oo n!} \, 
     {{\p X^{\mu_1 ...\mu_n}}\oo {\p {\tilde \xi}}}
     {{\p X_{\mu_1 ...\mu_n}}\oo {\p {\tilde \xi}}} - \lambda^2 = 0 
     \lbl{2.46} \\
     \delta X^{\mu_1 ...\mu_n} &:& \quad {\dd \oo {\dd {\tilde \xi}}}
     \left ( {1\oo \lambda} \, {{\p X^{\mu_1 ...\mu_n}}\oo {\p {\tilde \xi}}}
     \right ) = 0 \lbl{2.47}
\ear
These are the equations of motion for the $n$-area variables.

Inserting eq.(\ref{2.46}) into (\ref{2.44}) we obtain the action which 
is a
functional of $X^{\mu_1 ...\mu_n}$ solely:
\be
    I[X^{\mu_1 ...\mu_n}] = \kappa \int \dd {\tilde \xi} \, 
    \left ( {1\oo n!} \,
    {{\p X^{\mu_1 ...\mu_n}}\oo {\p {\tilde \xi}}}
    {{\p X_{\mu_1 ...\mu_n}}\oo {\p {\tilde \xi}}} \right )^{1/2}
\lbl{2.48}
\ee
The latter action has the same form as the action for a worldline of a
relativistic particle.
Factor $1/n!$ in the latter action disappears, if
we order indices according to $\mu_1 < \mu_2 < ... < \mu_n$.

\subsection{More general objects (generalization to $C$-space)}

So far we have considered branes described by coordinate functions 
of the type
$X^{\mu} (\xi),~X^{\mu_1 \mu_2} (\xi^{a_1 a_2})$,...,
or
$X^{\mu_1 ...\mu_n} (\xi^{a_1 ... a_n})$ that represent a mapping of a
worldsheet loop into a target space loop of the same dimensionality.:
    $$ (n-1)\mbox{-loop on} \; V_n 
    \longrightarrow (n-1)\mbox{-loop in } V_N \, .$$
Let us now extend the theory and consider `` mixed'' mappings
     $$ (r-1)\mbox{-loop on} \; V_n 
     \longrightarrow (R-1)\mbox{-loop in } V_N \, .$$
where $r$, in general, is different from $R$.
So we arrive at a more general extended object which is described by a mapping
\ci{PavsicBook,CastroPavsicReview}
\bear
      &&\xi^{a_1 ... a_r} \rightarrow
      x^{\mu_1 ... \mu_R} = X^{\mu_1 ... \mu_R} (\xi, \xi^a, 
      \xi^{a_1 a_2},..., \xi^{a_1 ... a_s},..., \xi^{a_1...a_n}) \; ,
      \nonumber \\
      &&\hs{6cm} \qquad 0 \le R \le N \; , \quad 0 \le r \le n < N
\lbl{2.59}
\ear
In the compact notation we set
\bear
       &&X^M \equiv X^{\mu_1 ... \mu_R} \; , 
       \qquad \mu_1 < \mu_2 < ... < \mu_R
       \nonumber \\
       &&\xi^A \equiv \xi^{a_1...a_r} \; , \qquad a_1 < a_2 < ... < a_r 
       \nonumber
\ear       
and write the mapping (\ref{2.59}) as
\be
       \xi^A \rightarrow x^M = X^M (\xi^A)
\lbl{2.60}
\ee
This is the parametric equation of our generalized extended object. 
Such object
lives in a target space which is now generalized to
{\it Clifford space} (shortly $C$-space).
The worldsheet associated with the extended 
object is also generalized to
a Clifford space. In the following we will explain this in more detail.

In eq.\,(\ref{2.59}) or (\ref{2.60})we have a generalization of the usual
relation
\be
    \xi^a \rightarrow x^\mu = X^\mu (\xi^a)\, ,~~~a=1,2,...,n; ~~~
    \mu = 0,1,2,..., N-1
\lbl{L1}
\ee
that describes an $n$-dimensional surface, called worldsheet or
world manifold, $V_n$, embedded in
$N$-dimensional target space $V_N$. In eq.\,(\ref{L1}) the space 
$\mathbb{R}^n$ of
parameters $\xi^a$ is isomorphic to an $n$-dimensional vector space
${\bV}_n$, spanned by an orthonormal basis $\lbrace h_a \rbrace$.
The {\it vector space} ${\bV}_n$ should not be confused with the worldsheet
$V_n$, which is a {\it manifold} (embedded in a higher dimensional
manifold $V_N$).

Instead of ${\bV}_n$ we can consider the corresponding Clifford algebra
${\boldsymbol C}_n$ which is itself a vector space. Amongst its elements
are $r$-vectors associated with
$(r-1)$-loops, $r=0,1,2,..., n$. A generic object
is a superposition of $r$-vectors for different grades, and it is
described by a Clifford number, a polyvector,
$\xi^{a_1 ...a_r} h_{a_1} \wg ... \wg h_{a_r} \in {\boldsymbol C}_n$

Our objects are now {\it extended events} ${\cal E}$ \ci{PavsicKaluzaLong},
superpositions of $(r-1)$-loops, to which we assign a set of
$2^n$ parameters (coordinates) $\xi^A \equiv \xi^{a_1 ... a_r}$,
$r=0,1,2,...,n$ according to the mapping
\be
     {\cal E} \rightarrow \xi^A ({\cal E})
\lbl{L2}
\ee
The assignment is arbitrary. We may choose an object ${\cal E}_0$ to which
we assign coordinates $\xi^A ({\cal E}) = 0$. This is a coordinate origin.
Choosing an origin ${\cal E}_0$, the polyvectors $\xi^A h_A$ pointing
from ${\cal E}_0$
to any ${\cal E}$ are in one-to-one correspondence with
extended events ${\cal E}$. The space of extended events is then
isomorphic to Clifford algebra ${\boldsymbol C}_n$, and the latter algebra,
in turn, is isomorphic to the space of parameter
$\lbrace \xi^A \rbrace = {\mathbb R}^{2^n}$. Therefore we will
speak about ${\boldsymbol C}_n$ as the 
{\it parametric Clifford algebra} or {\it parametric polyvector space}.

The parametric space ${\boldsymbol C}_n$  is by definition a
{\it (poly)vector} space,
spanned by a basis $h_{a_1} \wg ... \wg h_{a_r}$, $r=0,1,2,...,n$,
formed by the {\it orthonormal basis} $\lbrace h_a \rbrace $, $a=1,2,...,n$.
This implies that ${\boldsymbol C}_n$ is a metric space, but its metric is
just formal, without any physical content. Now let us consider the
mapping (\ref{2.60}) from ${\boldsymbol C}_n$ into a Clifford space $C_{V_N}$
generated by a basis $\gam_\mu\, , ~\mu=0,1,2,...,N-1$, with $N>n$.
So we obtain a generalized, $2^n$-dimensional surface
$C_{V_n}$ embedded in a target Clifford space $C_{V_N}$. The surface
$C_{V_n}$ generalizes the notion of worldsheet $V_n$ to Clifford space,
i.e., a manifold such that any of its tangents spaces is a Clifford
algebra. If we consider only the intrinsics propertires of $C_{V_n}$
(i.e., if we ``forget" about its embedding into a higher dimensional
Clifford space $C_{V_N}$), then we can simply denote it as $C_n$.

Bellow we sumarize our notation of various spaces:

\begin{description}

\item[${\boldsymbol V}_n$] \ {\it Parametric vector space}, with 
an orthonormal basis $\lbrace h_a \rbrace$, $a=1,2,...,n$  and elements 
$\xi^a h_a \in {\boldsymbol V}_n$.
It is isomorphic to ${\mathbb R}^n$, the space of
parameters $\xi^a$.

\item[$V_n$] \ {\it Manifold}, either flat or curved. It is a space
of points (events) ${\cal P}$. With every point ${\cal P} \in V_n$ we
associate a set of $n$ parameters (coordinates)
$\xi^a ({\cal P}) \equiv \xi^a \in {\mathbb R}^n$.
Coordinate basis vectors are $e^a$, whilst orthonormal basis vectors
are $e^{\bfs a}$.

\item[${\boldsymbol C}_n$] \ {\it Parametric Clifford algebra} 
of ${\boldsymbol V}_n$, called
also {\it parametric polyvector space},
with basis
$\lbrace h_A \rbrace \equiv \lbrace h_{a_1} \wg ... \wg h_{a_r} \rbrace,
~ r = 0,1,2,...,n$ and elements $\xi^A h_A \in {\boldsymbol C}_n$, called
{\it polyvectors}. It is isomorphic to ${\mathbb R}^{2^n}$,
the space of parameters $\xi^A$.

\item[$C_n$] \ {\it Clifford manifold}, or {\it Clifford space},
either flat or curved. It is a space of points that are perceived in
a subspace $V_n$ as {\it extended events} ${\cal E}$. With every
${\cal E} \in C_n$ we associate a set of $2^n$ parameters
$\xi^A ({\cal E}) \equiv \xi^a \equiv \xi^{a_1 ... a_r}
\in {\mathbb R}^{2^n},~r=0,1,2,...,n$. Coordinate basis elements
are $e_A \equiv e_{a_1 ...a_r}$; orthonormal basis
elements are $e_{\bfs A} \equiv e_{{\bfs a}_1...{\bfs a}_n},~r=0,1,2,...,n$.
In particular $C_n$ can be considered as being embedded in a higher
dimensional Clifford space; then it is denoted as $C_{V_n}$.

\item[$C_{V_n}$] {\it Generalized worldsheet}, a Clifford space
embedded in a target Clifford space. Its coordinate basis elements
are $e^A \equiv e_{a_1 ...a_r}$; orthonormal basis
elements are $e_{\bfs A} \equiv e_{{\bfs a}_1...{\bfs a}_n},~r=0,1,2,...,n$.

\item[$C_{V_N}$] {\it Target Clifford manifold}, or {\it target
Clifford space}, either flat or curved. It is a space of points
that are perceived in a subspace $V_N$ as extended events ${\cal E}$.
With every ${\cal E} \in C_{V_N}$ we associate a set of $2^N$ coordinates
$x^M ({\cal E}) \equiv x^M \equiv x^{\mu_1 ... \mu_R},~R=0,1,2,...N$.
Coordinate basis elements are $\gam_M \equiv \gam_{\mu_1 ... \mu_R}$;
orthonormal basis elements are $\gam_{\bfs M} \equiv \gam_{{\bfs \mu}_1...
{\bfs \mu}_R}$. Instead of $C_{V_N}$ we can use simply notation
$C_N$.

\end{description}

We follow the rule that bold symbols are used for vector spaces, whilst
light symbols are used for manifolds. Since Clifford algebras also
are vector spaces, they are denoted by bold symbols, whereas the
corresponding Clifford spaces (manifolds of points representing extended
events) are denoted by light symbols. By such notation we have attempted
to simplify distinction among all those various spaces that occur
in our theory of generalized branes. For these purely physical reasone
we have thus, to certain extent,
deviated from the standard notation
used in mathematics. 

An infinitesimal (polyvector) $\dd X \in C_{V_N}$, joining two points
on the surface $C_{V_n}$ can be written as
\be
    \dd X = \dd X^M \gam_M = \dd \xi^A \p_A X^M \gam_M = \dd \xi^A e_A
\lbl{L3}
\ee
where
\be
     e_A = \p_A X^M \gam_M 
\lbl{L4}
\ee
These are induced basis tangent (polyvectors) on $C_{V_n}$.

At every point of the {\it flat} target $C$-space $C_{V_N}$ there exists
a basis 
\be
     \lbrace \gamma^M \rbrace = \lbrace {\bf 1}, \gamma^\mu, 
     \gamma^{\mu_1 \mu_2}, ..., \gamma^{\mu_1 ... \mu_N} \rbrace
\lbl{2.61}
\ee
given in terms of
$2^N$ Clifford numbers
\be
      \gamma^M \equiv \gamma^{\mu_1 ... \mu_r} \equiv \gamma^{\mu_1} \wedge
      \gamma^{\mu_2} \wedge ... \wedge \gamma^{\mu_R} \; , \quad
      \mu_1 < \mu_2 < ... \mu_R
\; , ~~~ R = 0,1,2,... N
\lbl{2.61a}
\ee

At every point $\xi \in C_{V_n}$
on the brane's worldsheet $C$-space $C_{V_n}$,
which in general is {\it curved},
there exist a basis given in terms of
$2^n$ Clifford
numbers\footnote{We will use `basis' and `frame'
as synonyms. In order to simplify notation and wording we will be
sloppy in distinguishing objects from the corresponding fields,
e.g., (poly)vectors from (poly)vector fields, frames from frame
fields, etc. From the context it should not be difficult to
understand which ones we have in mind.}
\be
     \lbrace e^A \rbrace = \lbrace e, e^a, e^{a_1 a_2}, ...,
e^{a_1 ...a_n} 
     \rbrace
\lbl{2.62}
\ee
that  span a tangent space $T_\xi (C_{V_n})$.
At a particular point $\xi_0 \in C_{V_n}$ it may hold
\be
     e^A \equiv e^{a_1 ... a_r} \equiv e^{a_1} \wedge e^{a_2}
     \wedge e^{a_r} \; , \quad a_1 < a_2 < ... < a_r 
     \; , \quad r=0,1,2,...,n
\lbl{2.62a}
\ee
That is, at that particular point the basis polyvectors on $C_{V_n}$ 
are given as wedge products of basis vectors $e^a$.
The above property (\ref{2.62a}) cannot hold globally on a {\it curved}
$C_{V_n}$.
At points different from $\xi_0$, basis polyvectors are in general
superpositions of $e^{a_1} \wedge e^{a_2} \wedge ...
\wedge e^{a_s}$,
$s=0,1,2,...,n$.

To sum up, every Clifford number in the target $C$-space can be expanded
in terms
of 
$\gamma^M$, and every Clifford number on the worldsheet $C$-space can
be expanded in terms of $e^A$. Such numbers are also called {\it Clifford
aggregates} or {\it polyvectors}. They are superpositions $r$-vectors,
the objects of definite grade that we call multivectors.
This description automatically contains {\it spinors} which are just
members of the left of right ideal of Clifford algebra \ci{IdealSpin}.
 
Metric of ${C}_{V_N}$ is $G_{MN} = \gam_M^{\ddg} * \gam_N$,
whilst  metric of ${C}_{V_n}$ is $G_{AB}=e_A^\ddg *e_B$. Here `$*$'
denotes the scalar
product of two Clifford numbers $A$ and $B$
\be
      A*B = \langle A B \rangle_0
\lbl{2.62b}
\ee      
    
Let us now define the object $V$ which is a polyvector in target space
and on the worldsheet:
 \be
     V= e^A {{\p X^M}\oo {\p \xi^A}} \, \gamma_M =
     \sum_{r=0}^N \sum_{s=0}^n {1\oo {r! s!}} e^{a_1 ...a_s}
     \, {{\p X^{\mu_1 ... \mu_r}}\oo {\p \xi^{a_1 ... a_s}}} \, 
     \gamma_{\mu_1 ...\mu_r} 
\lbl{2.63}
\ee
In the right hand side expression we impose no restriction on the 
indices 
$\mu_1$, $\mu_2,...,\mu_r$ and $a_1$, $a_2,...,a_s$,
therefore, in order to prevent 
multiple counting of equivalent terms, factors $1/r!$ and $1/s!$ are introduced.

It is illustrative to express eq.(\ref{2.63}) in a more explicit form by
employing the notation (\ref{2.61})--(\ref{2.62a}) and by writing
\bear
     &&X^M = (\Omega, X^\mu, X^{\mu_1 \mu_2}, ..., X^{\mu_1 ... \mu_N})
     \lbl{2.64} \\
     &&\xi^A = (\xi, \xi^a, \xi^{a_1 a_2}, ..., \xi^{a_1 ... a_n})
     \lbl{2.65}
\ear
We obtain
\bear
   V &=& \left ( {{\p \Omega}\oo {\p \xi}} + e^a \, {{\p \Omega}\oo {\p
   \xi^a}} + {1\oo 2!} e^{a_1 a_2}{{\p \Omega}\oo {\p \xi^{a_1 a_2}}} + ...
   {1\oo n!} e^{a_1 ...a_n}{{\p \Omega}\oo {\p \xi^{a_1 ... a_n}}} 
   \right ) 
   {\underline 1}
   \nonumber \\
   &&+ \left ( {{\p X^\mu}\oo {\p \xi}} + e^a \, {{\p X^\mu}\oo {\p
   \xi^a}} + {1\oo 2!} e^{a_1 a_2}{{\p X^\mu}\oo {\p \xi^{a_1 a_2}}} + ...
   {1\oo n!} e^{a_1 ...a_n}{{\p X^\mu}\oo {\p \xi^{a_1 ... a_n}}} 
   \right ) 
   \gamma_\mu
   \nonumber \\
   &&+ {1\oo 2!} \left ( {{\p X^{\mu_1 \mu_2}}\oo {\p \xi}} + e^a \, 
   {{\p X^{\mu_1 \mu_2}}\oo {\p
   \xi^a}} + 
   {1\oo 2!} e^{a_1 a_2}{{\p X^{\mu_1 \mu_2}}\oo {\p \xi^{a_1 a_2}}} 
   + ...
   {1\oo n!} e^{a_1 ...a_n}{{\p X^{\mu_1 \mu_2}}\oo 
   {\p \xi^{a_1 ... a_n}}} 
   \right ) 
\gamma_{\mu_1 \mu_2} \lbl{2.66} \\
   &&+ \nonumber \\
   &&\vdots \nonumber \\
   &&+ {1\oo N!} \left ( {{\p X^{\mu_1... \mu_N}}\oo {\p \xi}} + e^a \, 
   {{\p X^{\mu_1 ...\mu_N}}\oo {\p
   \xi^a}} + {1\oo 2!} e^{a_1 a_2}
   {{\p X^{\mu_1 ... \mu_N}}\oo {\p \xi^{a_1 a_2}}} 
   + ...
   {1\oo n!} e^{a_1 ...a_n}{{\p X^{\mu_1 ...\mu_N}}\oo 
   {\p \xi^{a_1 ... a_n}}} 
   \right ) 
   \gamma_{\mu_1 ... \mu_N}
   \nonumber
\ear

One might ask how such a generalized extended object, described by
eq.\,(\ref{2.60}), that sweep a Clifford worldsheet
$C_{V_n}$, embedded in a Clifford space $C_{V_N}$, looks like.
Here our perceptive system again shows
its shortcomings, like in the case of figuring out how higher
dimensional objects look like. We are able to draw pictures of
projections of an object onto 3 or 2-dimensional Euclidean
space, and that is, more or less, all. But on the other hand, we are
able to do algebra, and the algebra is interpreted as geometric algebra.
So we have to content us by our ability to control the situation
algebraically, and assume that there is a mapping between algebraic
and geometric objects. The latter objects are associated with
physical objects, such as, e.g., the generalized extended objects,
that incorporate branes of various dimensionalities.

Let the action describing the dynamics of a generalized extended object,
shortly, a generalized brane, 
be described by the embedding functions $X^M (\xi^A)$ be
\be
    I[X^M, e^A] = {T\oo 2} \int \dd^{d_C} {\tilde \xi} \, |E| \,
    \left [ \left ( e^A \, {{\p X^M}\oo {\p \xi^A}} 
    \gamma_M \right )^{\ddg}
    * \left ( e^B \, {{\p X^N}\oo {\p \xi^B}} \gamma_N \right ) + 2 - 
    d_C  \right ]
\lbl{2.67}
\ee
The latter action has a similar form as the action (\ref{2.31}). But, since
the indices $M$, $N$,  and $A$, $B$ run over the full basis (\ref{2.61})
and (\ref{2.62}) of the corresponding $C$-spaces, the action (\ref{2.67})
is more general than (\ref{2.31}).

Here the measure $\dd^{d_C} \xi |E|$ is the volume element in the
worldsheet $C$-space $C_{V_n}$
(whose dimension is $2^{d_C}$). It is equal
to the product
\be
   \dd^{d_C} \xi \, |E| \equiv |E| \prod_{A_i} \, \dd \xi^{A_i}
   =\dd \xi \prod_{a_1}\dd \xi^a \prod_{a_1 < a_2}
    \dd \xi^{a_1 a_2} ... \prod_{a_1 < ...< a_n}  
    \dd \xi^{a_1 ... a_n}] |E|
\lbl{2.78}
\ee 
By $|E|$ we denote the square root of the determinant of the worldsheet
$C$-space metric which is given by the scalar product
\be
    G^{AB} = (e^A )^{\ddagger} * e^B = \langle (e^A )^{\ddagger} e^B \rangle_0
\lbl{2.79}
\ee
where $\langle ~ \rangle_0$ denotes the scalar part. Explicitly,
\be
    |E| = \sqrt{|G|} \; , \quad  G \equiv {\rm det} \, G_{AB} = 
    {1\oo d_C !} \,
    \epsilon^{A_1 ... A_{d_C}} \, \epsilon^{B_1 ... B_{d_C}} \, G_{A_1 B_1} ...
    G_{A_{d_C} B_{d_C}}
\lbl{2.80a}
\ee

The action (\ref{2.67}) is a functional of $X^M$ and $e^A$, and is a
$C$-space generalization of the action (\ref{2.31}) which is a functional
of the worldsheet embedding functions $X^\mu$ and basis vectors
$e^a$.

An action which is
classically equivalent to (\ref{2.67}) is a functional of $X^M$ and $G^{AB}$:
\be
    I[X^M , G^{AB}] = {T\oo 2} \int \dd^{d_C} \xi \, 
    \sqrt{|G|} \, (
    G^{AB} \, \p_A X^M \, \p_B X_M + 2 - d_C )
\lbl{2.80}
\ee
where $\p_B X_M = G_{MN}\, \p_B X^N \, ~, G_{MN} = \gam_M^\ddg * \gam_N$.
In eq.(\ref{2.80}) we have a $C$-space generalization of the well known
Howe--Tucker action \ci{Howe-Tucker}.

Variation of the action (\ref{2.80}) with respect to $X^M$ and $G^{AB}$ gives
the equations of motion of our $C$-space extended object:
\bear
       \delta X^M  &:& \qquad {1\oo {\sqrt{|G|}}} \, 
       \p_A (\sqrt{|G|} \, 
       \p^A X^M ) = 0 \lbl{2.91} \\
     \delta G^{AB} &:& \qquad G_{AB} = \p_A X^M \, \p_B X_M \lbl{2.92}
\ear
Taking into account that $G_{AB} G^{AB} = d_C = 2^n$ and inserting eq. 
(\ref{2.92})
Into eq. (\ref{2.80}) we obtain the action integral
\be
    I[X^M] = T \int \dd^{d_C}  \xi \, \sqrt{{\rm det} \, \p_A X^M \, \p_B X_M}
\lbl{2.93}
\ee
which is the volume of the $C$-space worldsheet. The latter action
contains the usual $p$-branes, including point particles, as special
cases.

Our action (\ref{2.80}), or equivalently (\ref{2.93}), is invariant under
reparametrizations of coordinates $\xi^A,~A=1,2,...,2^n$.
As a consequence, there are $2^n$ primary constraints. So we are
free to choose $2^n$ relations among our dynamical variables
$X^M (\xi^A)$, and thus fix a gauge (a parametrization). For one
of those relations we can choose, for instance,
\be
      \p_{a_1 ...a_r} X^{\mu_1 ...\mu_r} =
   \p_{[a_1} X^{\mu_1} .... \p_{a_r]} X^{\mu_r}
\lbl{2.94}
\ee
That is, the above relation is just one of gauge fixing
relations. It will be used in Sec.\,5.3, where we will consider
the coupling of our generalized branes to external fields.

\section{On the relativity in Clifford space} 

The discussion of previous sections has led us to the conclusion that
the space in which our extended objects live is {\it Clifford space},
shortly $C$-space, denoted $C_{V_N}$ or $C_N$.
A point in $C$-space is
described by the coordinates
$x^M = (\Omega, x^\mu, x^{\mu_1 \mu_2}, ..., x^{\mu_1...\mu_N})$ which
together with the basis elements $\gamma^M = (1,\gamma_\mu, \gamma_{\mu_1
\mu_2},...,\gamma_{\mu_1 ... \mu_N})$ form a coordinate polyvector\footnote{
In flat $C$-space it makes sense to consider a polyvector joining
the coordinate origin ${\cal E}_0$ with coordinates $x^M ({\cal E}_0) = 0$
and a point ${\cal E}$ with coordinates $x^M ({\cal E}) \equiv x^M$,
where ${\cal E}_0,~ {\cal E} \in C_{V_N}$.}
\be
      X = x^M \gamma_M = {1 \oo r!} \sum_{r=0}^N x^{\mu_1 ...\mu_r}
      \gamma_{\mu_1 ... \mu_r}
\lbl{4.1}
\ee

From the point of view of $2^N$-dimensional $C$-space,
$x^{\mu_1 ... \mu_r}$, $r=1,...,N$, are coordinates of a point, whilst
from the point of view of the underlying Minkowski space $V_N$, these
are the $r$-area ($r$-volume) variables.

The infinitesimal polyvector connecting two infinitesimally separated
points of $C$-space is
\be
    \dd X  = \dd x^M \gamma_M = {1 \oo r!} \sum_{r=0}^N \dd x^{\mu_1 ...\mu_r}
      \gamma_{\mu_1 ... \mu_r} \equiv \dd x^M \gamma_M
\lbl{4.1a}
\ee
The square of the distance between these points is given by the scalar
product
\be
    |\dd X|^2 \equiv \dd X^{\ddg} * \dd X = \dd x^M \dd x^N G_{MN}
    = \dd x^M \dd x_N
\lbl{4.1b}
\ee
where $G_{MN}$ is the metric of $C$-space\footnote{
A reason why we define the $C$-space metric by employing the reversion
is in the consistency between $G_{MN}$, its inverse $G^{MN}$, and the relations
(\ref{4.3}),(\ref{4.4}) (in which the indices $\mu_i,~\nu_j,...$ are lowered
and raised by the ordinary 4-dimensional metric $g_{\mu_i \nu_j}$ and
its inverse $g^{\mu_i \nu_j}$). For more details see ref. \ci{PavsicSaasFee}.}
:
\be
     G_{MN} = {\gamma^{\ddagger}}_M * \gamma_N = {\gamma^{\ddagger}}_{\mu_1
     ...\mu_r} * \gamma_{\nu_1 ...\nu_s}
\lbl{4.2}
\ee  

In particular,
$C$-space $C_{V_N}$ can be flat. This means that the curvature of
the connection 
on flat $C_{V_N}$ vanishes (see sec.\,5.1). In such a case one can find
a coordinates system on $C_{V_N}$ such that the metric $G_{MN}$
is diagonal.
 
Explicitly, for different choices of the indices $M$ and $N$ we have:
\bear
    G_{\mu \nu} &=& \gamma_\mu \cdot \gamma_\nu = g_{\mu \nu} = 
    \eta_{\mu \nu} \nonumber \\
    G_{[\mu_1 \mu_2][\nu_1 \nu_2]} &=& \gamma_{\mu _1 \mu_2}^{\ddg}
    *\gamma_{\nu_1 \nu_2} = (\gamma_{\mu_2} \wedge \gamma_{\mu_1})\cdot
    (\gamma_{\nu_1} \wedge \gamma_{\nu_2}) 
    =  \begin{vmatrix} g_{\mu_1 \nu_1} & g_{\mu_1 \nu_2} \\
                g_{\mu_2 \nu_1} & g_{\mu_2 \nu_2}  
                \end{vmatrix} \nonumber \\
    G_{[\mu_1 \mu_2 \mu_3][\nu_1 \nu_2 \nu_3]} &=& 
    \gamma_{\mu _1 \mu_2 \mu_3}^{\ddg}
    *\gamma_{\nu_1 \nu_2 \nu_3} = (\gamma_{\mu_3} 
    \wedge \gamma_{\mu_2} \wedge 
    \gamma_{\mu_1})\cdot
    (\gamma_{\nu_1} \wedge \gamma_{\nu_2} \wedge \gamma_{\nu_3}) = 
      \begin{vmatrix} g_{\mu_1 \nu_1} & g_{\mu_1 \nu_2} & 
              g_{\mu_1 \nu_3} \\
              g_{\mu_2 \nu_1} & g_{\mu_2 \nu_2} & g_{\mu_2 \nu_3} \cr
              g_{\mu_3 \nu_1} & g_{\mu_3 \nu_2} & g_{\mu_3 \nu_3}
              \end{vmatrix}
     \nonumber \\
     G_{\mu [\nu_1 \nu_2]} &=& \gamma_\mu * \gamma_{\nu_1 \nu_2} =
     \langle \gamma_\mu (\gamma_{\nu_1} \wedge \gamma_{\nu_2} 
     \rangle_0 = 0   
\lbl{4.3}
\ear

In general we have
\be
    G_{[\mu_1 ... \mu_r][\nu_1 ... \nu_r]} =
    (\gamma_{\mu_r} \wedge ... \wedge \gamma_{\mu_1}) \cdot
    (\gamma_{\nu_1} \wedge ... \wedge \gamma_{\nu_r}) 
    = \mbox{\rm det} \, g_{\mu_i \nu_j} \; , \quad i,j = 1,2,...,r
\lbl{4.4}
\ee
when $r=s$, and
\be
     G_{[\mu_1 ... \mu_r][\nu_1 ... \nu_s]} = 0
\lbl{4.5}
\ee
when $r \neq s$.

Taking into account the explicit expressions for metric 
(\ref{4.3})--(\ref{4.5}), the quadratic form (\ref{4.1b}) reads
\bear
     |\dd X|^2 &=& {1\oo r!} \sum_{r=0}^N \dd x^{\mu_1 ... \mu_r}
     \, \dd x_{\mu_1 ... \mu_r} \nonumber \\
     &=&
     \dd \Omega^2 + \dd x^\mu \dd x_{\mu} + {1\oo 2!} \dd x^{\mu_1 \mu_2}
     \dd x_{\mu_1 \mu_2} + ... + {1 \oo N!} \dd x^{\mu_1 ... \mu_N}
     \dd x_{\mu_1 ... \mu_N}
\lbl{4.6}
\ear
In the latter expression only the factor $1/r!$ remains, since the other
factor is canceled by $r!$ coming from the determinant (\ref{4.4}).
Indices $\mu_1,~\mu_2,...$ are lowered and raised by Minkowski metric
$\eta_{\mu \nu}$ and its inverse $\eta^{\mu \nu}$. 

In the 16-dimensional Clifford space ${C}_{M_4}$ of the 4-dimensional
Minkowski spacetime the polyvector (\ref{4.1a}) and its square
(\ref{4.1b}) can be rewritten as
\be
  \dd X = \dd \Omega + \dd x^\mu \gamma_\mu + {1\oo 2} \dd x^{\mu \nu}
  \gamma_\mu \wedge \gamma_\nu + \dd {\tilde x}^\mu \gamma_5 \gamma_\mu
  + \dd {\tilde \Omega} \gamma_5
\lbl{4.7a}
\ee
\be
   |\dd X |^2 = \dd \Omega^2 + \dd x^\mu \dd x_\mu + {1\oo 2} 
   \dd x^{\mu \nu} \dd x_{\mu \nu} - \dd {\tilde x}^\mu {\tilde x}_\mu
   - \dd {\tilde s}^2
\lbl{4.7b}
\ee
where
\be
     \dd {\tilde x}^\mu \equiv {1\oo 3!} \dd x^{\alpha \beta \rho}
     {\epsilon_{\alpha \beta \rho}}^\mu
\lbl{4.8}
\ee
\be
     \dd {\tilde \Omega} \equiv {1\oo 4!} \dd x^{\alpha \beta \rho \sigma}
     \epsilon_{\alpha \beta \rho \sigma}
\lbl{4.9}
\ee
The minus sign in the last two terms of the above quadratic form occurs
because in 4-dimensional spacetime with signature $(+---)$ we have
$\gamma_5^2 = (\gamma_0 \gamma_1 \gamma_2 \gamma_3) 
(\gamma_0 \gamma_1 \gamma_2 \gamma_3) = -1$, and also
$\gamma_5^{\ddg} \gamma_5 = (\gamma_3 \gamma_2 \gamma_1 \gamma_0)
(\gamma_0 \gamma_1 \gamma_2 \gamma_3) = -1$.

It is illustrative to write the quadratic form (line element) explicitly:
\bear
     |\dd X|^2 &=&\dd \Omega^2 + (\dd x^0)^2 - (\dd x^1)^2 - (\dd x^2)^2
     - (\dd x^3)^2 \nonumber \\
     &&- (\dd x^{01})^1 - (\dd x^{02})^2 - (\dd x^{03})^2 + (\dd x^{12})^2
     +(\dd x^{13})^2 + (\dd x^{23})^2 \nonumber \\
     &&- (\dd {\tilde x}^0)^2 + (\dd {\tilde x}^1)^2 + (\dd {\tilde x}^2)^2
     + (\dd {\tilde x}^3)^2 - \dd {\tilde \Omega}^2
\lbl{4.7c}
\ear
Here ${\tilde x}^0 = x^{123},~{\tilde x}^1 = x^{023},~{\tilde x}^2 =
x^{013},~ {\tilde x}^3 = x^{012},~{\tilde \Omega}= x^{0123}$.
The factor $1/2$ has disappeared from the term $(1/2) \dd x^{\mu \nu}
\dd x_{\mu \nu}$, since $(1/2) (\dd x^{01} x_{01} + \dd x^{10} x_{10})
= \dd x^{01} \dd x_{01} = - (\dd x^{01})^2$, etc..

By inspecting the quadratic form (\ref{4.7b}) we see
that it has 8 terms with {\it plus} and 8 terms with {\it minus} sign.
The group of transformations that leave the quadratic form (\ref{4.7b})
invariant is SO(8,8). These are pseudorotations in $C$-space and have
the role of generalized Lorentz transformations:
\be
    x'^M = {L^M}_N x^N
\lbl{4.10}
\ee
The transformations matrix satisfies the relation ${L^M}_J {L^N}_K G_{MN}
= G_{JK}$.

If interpreted actively (\ref{4.10}) are the transformations that transform
a point of $C$-space with coordinates $x^M$ into another point with
coordinates $x'^M$. In the passive interpretation the point remains the
same, but its coordinates $x^M$ with respect to a reference frame $S$
are transformed into
coordinates $x'^M$ with respect to a reference frame $S'$.

From the point of view of $C$-space, $x^M$ are coordinates of a {\it point}.
But from the point of view of the underlying Minkowski space,
$x^M$ are the $(p+1)$-vector (holographic) coordinates of $p$-loops associated
with an extended object \ci{PavsicArena}. In $C$-space, $p$-loops of
different dimensionalities $p$ (i.e., points, closed lines, closed
2-surfaces, closed 3-volumes) are all described on the same footing
\ci{Pezzaglia, Castro, PavsicTelAviv, PavsicArena,AuriliaFuzzy},
and can be transformed
into each other by transformations (\ref{4.10}). Pseudo rotations
in $C$-space have thus the role of polydimensional rotations in $M_4$.
A point can be transformed into a line, and in general an $(R-1)$-loop
into an $(R'-1)$-loop. So the very dimensionality of a loop can change under
a transformation. This means that, when observed in a given reference
frame $S$, a loop's dimensionality can change from $(R-1)$ to $(R'-1)$.
Alternatively, dimensionality of the same loop, when observed from
different reference frames $S$ and $S'$, can look different. In short,
dimensionality of a loop depends on the observer (associated with a
given reference frame). Such relativity of dimensionality of a loop
also explains why in the mapping (\ref{2.59}),(\ref{2.60}) the
dimensionality of a loop in the parameter space $\lbrace \xi^a \rbrace
\equiv {\mathbb R}^n$ is in general
different from the dimensionality of the same loop with respect to a reference
frame in the target space $V_N$.

$C$-space in essence encodes the zero modes of $p$-loop configurations,
since $p$-loops space is {\it infinite} dimensional whereas $C$-space
is finite dimensional. As already mentioned before, a $p$-loop
configuration, in general, can consist of many loops.

The construction with $C$-space coordinates, SO(8,8) symmetry and
the brane equations of motion (\ref{2.91}) reminds us of the constructions
considered in refs.\,\ci{DuffDual}, where extra coordinates were introduced
in order to make manifest the SO(n,n) symmetry of the duality
transformations for strings and branes.

\section{Curved Clifford space}

\subsection{General considerations}

In general, the worldsheet $V_n$ swept by a brane is {\it curved}. In
Sec.\,3.2 we have considered a concept of a more complicated, generalized
brane, whose (generalized) worldsheet ${C}_{V_n}$ was {\it curved Clifford
space}. The latter worldsheet ${C}_{V_n}$ was embedded in a
target space which was a {\it flat Clifford space}  ${C}_{V_N}$,
with the metric properties given in Sec.\, 4.

A next step is to consider {\it curved target Clifford space}
${C}_{V_N}$ (see refs.\,\ci{PavsicKaluza,PavsicKaluzaLong}). At every point
${\cal E} \in {C}_{V_N}$ we have a flat tangent Clifford space
$T_{\cal E} ({C}_{V_N})$ and an orthonormal basis of $2^N$
Clifford numbers
\be
    \lbrace \gam_{\bfs M} \rbrace=  \lbrace {\bf 1},\gam_{{\bfs \mu}_1},
    \gam_{{\bfs \mu}_1 {\bfs \mu}_2},..., 
    \gam_{{\bfs \mu}_1 {\bfs \mu}_2...  {\bfs \mu}_N} \rbrace
\lbl{5.1}
\ee
where
\be
   \gam_{{\bfs \mu}_1 {\bfs \mu}_2...  {\bfs \mu}_r} =
    \gam_{{\bfs \mu}_1} \wg \gam_{{\bfs \mu}_2}\wg...
    \wg \gam_{{\bfs \mu}_r} 
\lbl{5.2}
\ee

From an {\it orthonormal basis} $ \lbrace \gam_{\bfs M} \rbrace$
we can switch to a {\it coordinate basis}
\be
    \lbrace \gam_{ M} \rbrace = \lbrace \gam,\gam_{{ \mu}_1},
    \gam_{{ \mu}_1 {\mu}_2},..., 
    \gam_{{\mu}_1 {\mu}_2...  {\mu}_N} \rbrace
\lbl{5.3}
\ee
by means of the relation \ci{PavsicKaluza,PavsicKaluzaLong} 
\be
    \gam_M = {e_M}^{\bfs M} \gam_{\bfs M}
\lbl{5.4}
\ee
in which we have introduced a vielbein of curved Clifford space
${C}_{V_N}$, given by the scalar product $\gam_M^\ddg * \gam^{\bfs M}$.
Notice a distinction between bold and normal indices, used for two
different kinds of basis.
The coordinate basis Clifford numbers
$\gam_M =  \gam_{{\mu}_1 {\mu}_2...  {\mu}_r}$ in general are {\it not}
defined as a wedge product
$\gam_{\mu_1} \wg \gam_{\mu_2} \wg ... \wg \gam_{\mu_r}$. In particular,
$\gam_M$ can be multivectors of definite grade, i.e., defined as
a wedge product, but such property can hold only {\it locally} at a
given point ${\cal E}\in C_{V_N}$, and cannot be preserved globally at all
point ${\cal E}$ of our curved Clifford space. The relations
for $\gam_M$ and
the metric $G_{MN}$, discussed in sec.\,4, refer to flat $C$-space
and are no longer generally valid in curved $C$-space. At a fixed
point ${\cal E}\in {C}_{V_N}$ we can choose a coordinate system such
that $\gam_M = \gam_{\bfs M}$, and then the relations of Sec.\,4 refer
to flat $C$-space, spanned by $\gam_{\bfs M}$, i.e., the tangent space
$T_{\cal E} (C_{V_N})$.

The set $\lbrace \gam_M \rbrace$ of $2^n$ linearly independent
coordinate basis fields (which depend on coordinates $x^M$) will be called a
{\it coordinate frame field} in $C$-space.

The set $\lbrace \gam_{\bfs M} \rbrace$ of $2^n$ linearly independent
orthonormal basis fields (which also in general depend on $x^M$) will be
called {\it orthonormal frame field} in $C$-space.

Corresponding to each field $\gam_M$ we define a differential operator
---which we call {\it derivative}--- $\p_M$, whose action depends on the
quantity it acts on\footnote{
This operator is a generalization to curved $C$-space of the derivative
$\p_\mu$ which acts in an $n$-dimensional space $V_n$, and was defined by
Hestenes \ci{Hestenes} (who used a different symbol, namely $\Box_\mu$).}:

\ (i) $\p_M$ maps scalars $\phi$ into scalars
\be
    \p_M \phi = {{\p \phi}\oo {\p x^M}}
\lbl{A2.15}
\ee
Then $\p_M$ is just the ordinary {\it partial derivative}.

(ii) $\p_M$ maps Clifford numbers into Clifford numbers. In particular,
it maps a coordinate basis Clifford number $\gam_N$ into another Clifford
number which can, of course, be expressed as a linear combination of
$\gam_J$:
\be
     \p_M \gam_N = \Gam_{MN}^J \gam_J
\lbl{A2.16}
\ee
The above relation defines the connection $\Gam_{MN}^J$ for the coordinate
frame field $\lbrace \gam_M \rbrace$.

An analogous relation we have for the orthonormal frame field:
\be
    \p_M \gam_{\bfs M} = - {{\Omega_{\bfs M}}^{\bfs N}}_M \gam_{\bfs N}
\lbl{A2.17}
\ee
where ${{\Omega_{\bfs M}}^B}_M$ is the connection for the orthonormal frame field
$\lbrace \gam_{\bfs M} \rbrace$.

When the derivative $\p_M$ acts on a polyvector valued field
 $A=A^N \gam_N$ we obtain
\be
    \p_M (A^N \gam_N) = \p_M A^N \gam_N + A^N \p_M \gam_N =
    (\p_M A^N + \Gam_{MK}^N A^K) \gam_N \equiv \DD_M A^N \, \gam_N
\lbl{2.17a}
\ee
where $\DD_M A^N \equiv \p_M A^N + \Gam_{MK}^N A^K$ are components of
the covariant derivative in the coordinate basis, i.e., the `covariant
derivative' of the tensor analysis.

Here  $A^N$ are {\it scalar} components of $A$, and $\p_M A^N$ is just
the ordinary partial derivative with respect to $x^M$:
\be
    \p_M \equiv \left ( {\p \oo {\p s}},~{\p \oo {\p x^{\mu_1}}},~
    {\p\oo {\p x^{\mu_1 \mu_2}}}, ~{\p \oo {\p x^{\mu_1 ... \mu_n}}} 
    \right )
\lbl{A2.18}
\ee

The derivative $\p_M$ behaves as a {\it partial derivative} when acting on
scalars, and it defines a {\it connection} when acting on a basis
$\lbrace \gam_M \rbrace$. It has turned out very practical\footnote{
Especially when doing long calculation (which is usually the job of a
theoretical physicist) it is much easier and quicker to write $\p_M$
than $\Box_M, ~\nabla_M, ~D_{\gam_M}, ~\nabla_{\gam_M}$ which  all are
symbols used in the literature.}
to use
the easily writable symbol $\p_M$ which ---when acting on a
polyvector--- 
cannot be confused with partial derivative.
In ref.\,\ci{PavsicKaluzaLong}
we provided some arguments, why also conceptually it is more appropriate
to retain the same symbol $\p_M$, regardless of whether it acts on
scalar, vector, or generic polyvector fields.

The derivative $\p_M$ is defined with respect to a coordinate frame field
$\lbrace \gam_M \rbrace$ in $C$-space. We can define a more fundamental
derivative $\p$ by
\be
    \p = \gam^M \p_M
\lbl{A2.18a}
\ee
This is the {\it gradient} in $C$-space and it generalizes the ordinary
gradient $\gam^\mu \p_\mu$, $\mu = 0,1,2,...,n-1$, discussed by
Hestenes \ci{Hestenes}.

Besides the basis elements $\gam_M$ and $\gam_{\bfs M}$, we can define the
reciprocal elements $\gam^M,~\gam^{\bfs M}$ by the relations
\be
    (\gam^M)^\ddg * \gam_N = {\delta^M}_N \; , \quad
    (\gam^{\bfs M})^\ddg * \gam_{\bfs N} = {\delta^{\bfs M}}_{\bfs N}
\lbl{A2.19}
\ee

\paragraph{Curvature.} We define the curvature of $C$-space in the analogous
way as in the ordinary spacetime, namely by employing the commutator of
the derivatives \ci{Hestenes, PavsicBook, CastroPavsicHigher}.
Using eq.\,(\ref{A2.16}) we have
\be
   [\p_M,\p_N] \gam_J = {R_{MNJ}}^K \gam_K
\lbl{A2.19a}
\ee
where
\be
    {R_{MNJ}}^K= \p_M \Gam_{NJ}^K - \p_N \Gam_{MJ}^K + 
    \Gam_{NJ}^R \Gam_{MR}^K -
   \Gam_{MJ}^R \Gam_{NR}^K
\lbl{A2.20}
\ee
is the curvature of $C$-space. Using (\ref{A2.19a}) we can express the
curvature according to
\be
    (\gam^K)^\ddg * ([\p_M,\p_N] \gam_J) = {R_{MNJ}}^K
\lbl{A2.21}
\ee

An analogous relation we have if the commutator of the derivatives operates
on a orthonormal basis elements and use eq.\,(\ref{A2.17})

\subsection{On the dynamical curved $C$-space with sources}

We will assume that physics takes place in curved $C$-space. The latter space
is a generalization of curved spacetime. As in the ordinary general relativity
the metric $g_{\mu \nu}$ is a dynamical quantity, so in the
generalized general relativity the $C$-space metric  $G_{MN}$ is a dynamical
quantity. Instead of the usual point particles and branes the role of the
sources is now assumed by the generalized branes (which include the
generalized point particles) described by one of the classically
equivalent actions (\ref{2.67}),(\ref{2.80}) or (\ref{2.93}), which we
will denote by a common symbol $I_{\rm Brane}$.

The action thus contains a term which describes a $C$-space brane and
a kinetic term which describes the dynamics of the $C$-space itself:
\be
     I = I_{\rm Brane} + {1\oo {16 \pi G_C}} \int \dd^{2^n} x \, 
     \sqrt{|G|} \, R
\lbl{5.5}
\ee
Here $G_C$ is a constant (having the role of the ``gravitational''
constant in $C$-space), $G\equiv {\rm det}\, G_{MN}$ the determinant
of the $C$-space metric, and $R= R_{MNJK} G^{MJ} G^{NK}$ the curvature
scalar of $C$-space.

More details about how to proceed with and further generalize the
theory based on the action (\ref{5.5}) are to be found in 
\ci{PavsicBook}. The idea that the curved $C$-space can provide
a realization of Kaluza-Klein theory has been considered in
refs.\,\ci{PavsicParis,CastroKaluza,PavsicKaluza,PavsicKaluzaLong}.
In this
paper we will explore the brane part $I_{\rm Brane}$ which contains
the coupling of the $C$-space brane's embedding functions
$X^M (\xi^A)$ to the $C$-space metric $G_{MN}$.

\subsection{The generalized branes in curved $C$-space}

We will consider a generalized brane (a $C$-space brane) moving in a
fixed curved $C$-space background. We will assume that the action is
given by eq.\,(\ref{2.80}) in which there occurs the metric
$G_{MN}$ of the curved target $C$-space into which our generalized
brane is embedded.

Suppose that in the target $C$-space there exist isometries
described by $K$ Clifford numbers $k^\alpha = {k^\alpha}_M\, \gam^M ,~\alpha
= 1,2,...,K;~M= 1,2,...,16$, (generalized Killing ``vectors"),
 whose components satisfy the conditions
\be
   \DD_M {k^\alpha}_N +  \DD_N {k^\alpha}_M = 0
\lbl{5.4a}
\ee
where the covariant derivative $\DD_M A_N$ of components $A_M$
of an arbitrary polyvector $A$ is given in eq.\,(\ref{2.17a}).

Splitting the coordinate basis and orthonormal basis indices according to

~~~~~~~(i) {\it coordinate basis indices}: $M=(\mu,{\bar M}) \, ,\quad \mu
= 0,1,2,3 ;~{\bar M} = 4,5,...,16$

~~~~~~(ii) {\it orthonormal basis indices}: ${\bfs M} =({\bfs \mu},
{\bfs {\bar M}}) \; ,\quad {\bfs \mu}
= 0,1,2,3\, ;~{\bfs{ \bar M}} = 4,5,...,16$

the metric and vielbein can be written as 
\be
 G_{MN} = \begin{pmatrix}
            G_{\mu \nu} & G_{\mu {\bar M}}\\
            G_{{\bar M} \nu} & G_{{\bar M}{\bar N}}
            \end{pmatrix} \; , \quad 
    {e^{\bfs M}}_M = \begin{pmatrix}
            {e^{\bfs \mu}}_\mu & {e^{\bfs \mu}}_{\bar M} \\
            {e^{\bfs {\bar M}}}_\mu & {e^{\bfs{\bar M}}}_{\bar M}
            \end{pmatrix}             
\ee

Let us recall that that the vielbein according to eq.\,(\ref{4.5})
can be written as the scalar product of the $C$-space coordinate
and orthonormal basis elements:
\be
    {e_M}^{\bfs M} = \gam_M^\ddg *\gam^{\bfs M}
\lbl{5.6}
\ee
We can now assume that the orthonormal basis $\lbrace \gam_{\bfs M} \rbrace$
is chosen so that $\gam_{\bfs \mu}\, ,~{\bfs \mu} = 0,1,2,3$,
are tangent vectors to the spacetime (a subspace of $C$-space),
whilst the remaining basis elements $\gam_{\bfs {\bar M}}$, 
${\bar {\bfs M}} = 4,5,...,16$ are tangent to the ``internal"
part of $C$-space. Since the basis is orthogonal, we have
\be
    \gam_{\bfs \mu}^\ddg * \gam_{\bfs {\bar M}} = 0
\lbl{5.6a}
\ee
Next we assume that the coordinate basis $\lbrace \gam_M \rbrace$ is
chosen so that $\gam_\mu$, $~\mu = 0,1,2,3$ in general is
{\it not} tangent to $V_4$, whilst the remaining coordinate basis
elements $\gam_{\bar M}\, ,~{\bar M} = 4,5,...,16$, are tangent
to the ``internal" part of $C$-space. This means that
$\gam_\mu * \gam_{\bar M} \neq 0$, whilst
\be
    {\gam_{\bfs \mu}}^\ddg * \gam_{\bar M} = 0
\lbl{5.6b}
\ee
The latter equation can be written as 
$\eta_{{\bfs \mu}{\bfs \nu}}\,{\gam^{\bfs \nu}}^\ddg*\gam_{\bar M}=
 \eta_{{\bfs \mu}{\bfs \nu}}\,{e^{\bfs \nu}}_{\bar  M} = 0$.
Since $\eta_{{\bfs \mu}{\bfs \nu}}$ is diagonal, it follows that
\be
     {e^{\bfs \mu}}_{\bar M} = 0
\lbl{5.7}
\ee

Taking a coordinate system in which 
\be
     k^{\alpha \mu} = 0\, ,~~~~~~~   k^{\alpha {\bar M}} \neq 0
\lbl{5.7a}
\ee
the components ${e^{\bfs {\bar M}}}_\mu$
can be written in terms of the Killing vectors and gauge fields
${A_\mu}^\alpha (x^\mu)$:
\be
    {e^{\bfs {\bar M}}}_\mu = {e^{\bfs{\bar M}}}_{\bar M}
    \,k^{\alpha {\bar M}} {A_\mu}^\alpha \; ; \qquad ~~~~ 
    \p_{\bar M}{A_\mu}^\alpha = 0
\lbl{5.8}
\ee
For the ``mixed" components of the inverse vielbein we find
\be
     {e^\mu}_{\bfs {\bar M}} = 0
\lbl{5.9}
\ee
This follows directly from
\be
     {\gam^\mu}^\ddg * \gam_{\bar M} = 0 = {\gam^\mu}^\ddg
     ({e^{\bfs M}}_{\bar M} \,\gam _{\bfs M}) = {\gam^\mu}^\ddg
     *({e^{\bfs \mu}}_{\bar M} \gam_{\bfs \mu} +
      {e^{\bfs M}}_{\bar M}\,\gam_{\bfs {\bar M}} ) =
     {e^\mu}_{\bfs {\bar M}} {e^{\bfs{\bar M}}}_{\bar M}
\lbl{5.10}
\ee
where we have used eq.\,(\ref{5.7}). Since in general
${e^{\bfs{\bar M}}}_{\bar M} \neq 0$, it follows that
${e^\mu}_{\bfs {\bar M}}$ is equal to zero.

Using (\ref{5.4}),(\ref{5.7}) and (\ref{5.9}) and the above choice
of coordinates and local orthonormal frame we find that the quadratic
form can be written as the sum of the 4-dimensional quadratic
form and the part due  to the ``extra dimensions":
\be
    \dd X^\ddg * \dd X = (\dd x^M \gam_M)^\ddg * (\dd x^N \gam_N) = g_{\mu
    \nu} \, \dd x^\mu \dd x^\nu + \phi^{{\bar M}{\bar N}}
\, \dd x_{\bar M} \dd x_{\bar N}
\lbl{5.11}
\ee
where
\be
 g_{\mu \nu} = {e^{\bfs \mu}}_\mu
 {e^{\bfs \nu}}_\nu \gam_{\bfs \mu} \gam_{\bfs \nu}
 = G_{\mu \nu} - \phi^{{\bar M}{\bar N}}\,{k^\alpha}_{\bar M}
  {k^\beta}_{\bar N}{A_\mu}^\alpha {A_\nu}^\beta
 ~~~{\rm and}~~~~
 \phi^{{\bar M}{\bar N}} \equiv {e^{\bar M}}_{{\bfs {\bar M}}}
 {e^{\bar N}}_{{\bfs {\bar N}}}\, 
 \eta^{{\bfs {\bar M}}{\bfs {\bar N}} }
\lbl{5.12}
\ee
Here $G_{{\bar M}{\bar N}} \equiv \phi_{{\bar M}{\bar N}} =
{e^{\bfs {\bar M}}}_{\bar M}{e^{\bfs {\bar N}}}_{\bar N}
\,\eta_{{\bfs {\bar M}}{\bfs {\bar N}}}$, and $\phi^{{\bar M}{\bar N}}$
is the inverse of $\phi_{{\bar M}{\bar N}}$ in the ``internal" space.
Notice the validity of eqs.\,(\ref{5.7}) and (\ref{5.9}).

Let us now use the splitting (\ref{5.11}) in the brane action
(\ref{2.80}). We obtain
\be
    \p_A X^M \, \p_B X^N \,G_{MN} = \p_A X^\mu \p_B X^\nu \, g_{\mu \nu}
    + \p_A X_{\bar M} \p_B X_{\bar N} \, \phi^{{\bar M}{\bar N}}
\lbl{5.13}
\ee

The auxiliary variables $G_{AB}$ and the induced metric on the (generalized,
i.e., $C$-space) brane are related according to eq.\,(\ref{2.92}). 
Let us introduce new auxiliary variables $G'_{AB}$ and new brane tension
$T'$ according to
\be
    G_{AB} = G'_{AB} + \p_A X_{\bar M} \p_B X_{\bar N} \Phi^{{\bar M}{\bar
    N}}
\lbl{5.14}
\ee
\be
    T \sqrt{|G|} \, G^{AB} = T' \sqrt{|G'|} \, G'^{AB}
\lbl{5.15}
\ee
where $G^{AB}$ and $G'^{AB}$ are the inverse matrices of $G_{AB}$ and
$G'_{AB}$, respectively.

Using eqs.\,(\ref{5.14}),(\ref{5.15}) we have
\be
    T \sqrt{|G|} \, d_C =  T \sqrt{|G|} \, G^{AB} G_{AB} =
    T' \sqrt{|G'|} \, G'^{AB} (G'_{AB} + 
    \p_A X_{\bar M} \p_B X_{\bar N} \Phi^{{\bar M}{\bar N}})
\lbl{5.16}
\ee
\be
    T \sqrt{|G|} =\frac{ T \sqrt{|G|} \, d_C}{d_C} =
     \frac{T \sqrt{|G|} \, G^{AB} G_{AB}}{d_C} =
   \frac{ T' \sqrt{|G'|}}{d_C} \, G'^{AB} (G'_{AB} + 
    \p_A X_{\bar M} \p_B X_{\bar N} \Phi^{{\bar M}{\bar N}})
\lbl{5.17}
\ee
From eqs.\, (\ref{5.16}),(\ref{5.17}) we find that the extra term in
the brane action (\ref{2.80}) can be written
\be
     T \sqrt{|G|} \, (2 -d_C) = T' \sqrt{|G'|} \, (2 - d_C)
     + {P^A}_{\bar M} \,\p_A X_{\bar N}\,\Phi^{{\bar M}{\bar N}} 
     \left (\frac{2}{d_C} - 1 \right )
\lbl{5.18}
\ee
where we have written
\be
   T \sqrt{|G|} \, G^{AB} \p_A X_{\bar M} = 
    T' \sqrt{|G'|} \, G'^{AB} \p_A X_{\bar M} = {P^A}_{\bar M}
\lbl{5.19}
\ee
Here ${P^A}_{\bar M}$ are extra components of the brane canonical
momentum ${P^A}_{\bar M}= \p {\cal L}/\p \p_B X^M =
T \sqrt{|G|} \, G^{AB} \p_A X_{M}$.

Inserting eqs.\,(\ref{5.14}),(\ref{5.15}) and (\ref{5.18}) into the
brane action (\ref{2.80}) we obtain
\bear
   &&I_{\rm Brane} [X^\mu,G'^{AB}] = \frac{1}{2}\int \dd^{d_c}\xi\,
    T'\, \sqrt{|G'|}\, \left ( G'^{AB} \p_A X^\mu \p_B X^\nu g_{\mu \nu}
    + 2 - d_C \right )  \nonumber \\
     && ~~~~~~~~~~~~~~~~~~~~~~~~~+ \frac{1}{d_C} \int \dd^{d_c}\xi\,
     {P^A}_{\bar M} \p_A X_{\bar N} \Phi^{{\bar M}{\bar N}}
\lbl{5.20}
\ear

Variation of the latter action with respect to $G'^{AB}$ gives
\be
     G'_{AB} = \p_A X^{\mu} \p_B X^{\nu} g_{\mu \nu}
\lbl{5.21}
\ee
which is consistent with eqs.\,(\ref{2.92}),(\ref{5.13}),(\ref{5.14}).
Since the index $\mu$ runs over the embedding spacetime coordinates
$\mu = 0,1,2,..., N-1$ and $A,B = 1,2,..., d_C$ run over the coordinates
of the brane's  $C$-space worldsheet $C_{V_n}$
of dimension $d_C = 2^n$, we have
that the sytem of equations (\ref{5.21}) is determined, if $2^n \le
N$. In the case $N=4$ we have that $2^n \le 4$, which is satisfied for
$n\le 2$. In 4-dimensional spacetime the splitting described above works
for point particles ($n=1$) and strings ($n=2$). Higher dimensional branes,
that do not satisfy $2^n \le N$, of course, can also exist, but $G'_{AB}$
then cannot have the role of auxiliary variables, because the system
of equations (\ref{5.21}) is then underdetermined. In order to obtain
a determined system for auxiliary variables, a splitting of the indices
$A,~B$, analogous to that of $M,~N$ is then needed as well.

In eq.\,(\ref{5.20}) we have an action for a generalized $C$-space brane
worldsheet $C_{V_n}$ embedded in a higher dimensional curved manifold
(which in our case is the embedding $C$-space $C_{V_N}$).
In particular, if we
have just a usual point particle sweeping a 1-dimensional `worldsheet',
i.e., a worldline, embedded in a higher dimensional curved space, which
can be either a $C$-space or just simply a spacetime with extra dimensions,
then we have to take $d_C =1$, and the action (\ref{5.20}) becomes
\be
    I[X^\mu,\Lambda] = \frac{M}{2}\int \dd \tau \, \Lambda 
    \left (\frac{{\dot X}^{\mu} {\dot X}^\nu g_{\mu \nu} }{\Lambda^2} +1
    \right ) + \int \dd \tau P_{\bar M} {\dot X}_{\bar N} \Phi^{{\bar M}
    {\bar N}}
\lbl{5.22}
\ee
Above we have denoted 
\bear
&&G'_{AB} = G'_{11} \equiv \Lambda^2,~~~~ G'^{AB}=G'^{11}
\equiv \frac{1}{\Lambda^2},~~~~ T' \equiv M \nonumber \\
 &&\p_A X^M = \p_1 X^M \equiv 
\frac{\dd X^M}{\dd \tau} \equiv{\dot X}^M\nonumber \\
 && {P^A}_{\bar M} = {P^1}_{\bar M} \equiv P_{\bar M}
 \lbl{5.22a}
 \ear
The action (\ref{5.22}) can also be obtained by splitting
the following point particle action:
\be
 I[X^M,\lambda] = \frac{m}{2}\int \dd \tau \, \lambda 
    \left (\frac{{\dot X}^{M} {\dot X}^N G_{MN} }{\lambda^2} +1
    \right )
\lbl{5.23}
\ee
Variation of the latter action with respect to $\lambda$ gives
\be
    \lambda^2 = {\dot X}^M {\dot X}_M 
\lbl{ 5.24}
\ee
which can be split according to
\be
    {\dot X}^M {\dot X}_M = {\dot X}^{\mu} {\dot X}^\nu g_{\mu \nu}
    + {\dot X}_{\bar M} {\dot X}_{\bar N} \Phi^{{\bar M}{\bar N}}
\lbl{5.25}
\ee
Introducing a new auxiliary variable $\Lambda$ and new mass $M$
(a 4-dimensional mass) according to
\bear
   &&\lambda^2 = \Lambda^2  + 
    {\dot X}_{\bar M} {\dot X}_{\bar N} \Phi^{{\bar M}{\bar N}} \lbl{5.26}\\
   && \frac{m}{\lambda}= \frac{M}{\Lambda} \lbl{5.27}
\ear
and inserting eqs.\,(\ref{5.25})--(\ref{5.27}) into the point particle
action (\ref{5.23}) we obtain the split point particle action (\ref{5.22}),
where
\be
    P_{\bar M} = \frac{m}{\lambda}\, {\dot X}_{\bar M}
    = \frac{M}{\Lambda}\, {\dot X}_{\bar M}
\lbl{5.28}
\ee
So we have verified that the split generalized brane action\
(\ref{5.20}) includes a correct point particle action.
Relations (\ref{5.25})--(\ref{5.28}) above are point particle analog of
the brane relations (\ref{5.13})--(\ref{5.15}), (\ref{5.19}). 

Let us now return to the generalized brane action (\ref{5.20}). The
extra term can be written as
\bear
    {P^A}_{\bar M} \p_A X_{\bar N} \Phi^{{\bar M}{\bar N}} &=&
    {P^A}_{\bar M}\Phi^{{\bar M}{\bar N}} G_{{\bar N}J} \p_A X^J =
    {P^A}_{\bar M}\Phi^{{\bar M}{\bar N}} {A^\alpha}_J k_{{\bar N} \alpha}
    \p_A X^J \nonumber \\
    &=& {J^A}_\alpha {A^\alpha}_J \p_A X^J = {J^A}_\alpha ({A^\alpha}_\mu
    \p_A X^\mu + {A^\alpha}_{\bar M} \p_A X^{\bar M})
\lbl{5.29}
\ear
where
\be
    {J^A}_\alpha \equiv {P^A}_{\bar M} \Phi^{{\bar M}{\bar N}} 
    k_{{\bar N}\alpha} = {P^A}_{\bar M} k^{\bar M}_\alpha = 
    {P^A}_M k^M_\alpha
\lbl{5.29a}
\ee
are current densities which are conserved due to the presence of
isometries. The last step in eq.\,(\ref{5.29a}) holds because of
eq.\,(\ref{5.7a}). In eq.\,(\ref{5.29}) we have written the metric
components in terms of gauge potentials and Killing ``vectors"
\be
     G_{{\bar N}J} = {A^\alpha}_J k_{{\bar N}\alpha}
\lbl{5.30}
\ee
In particular,
\be
   G_{{\bar N}\mu} = {A^\alpha}_\mu k_{{\bar N} \alpha}
   \qquad~~~~~ {\rm if} ~~~ J=\mu
\lbl{5.31}
\ee
\be
   G_{{\bar N}{\bar M}} = {A^\alpha}_{\bar M} 
   k_{{\bar N}\alpha}\qquad~~~~~ {\rm if} ~~~ J={\bar M}
\lbl{5.32}
\ee
In eq.\,(\ref{5.31}) we have the well known relation
between gauge potentials, Killing ``vectors" and ``mixed" components
of the metric, the relation that was derived (see e.g., \ci{Luciani,Witten})
within the context of an ordinary higher dimensional spacetime manifold.
In eq.\,(\ref{5.32}) we have rewritten the ``internal"
space metric components in terms of gauge potentials ${A^\alpha}_{\bar M}$
and Killing ``vectors".

Inserting eq.\,(\ref{5.29}) into eq.\,(\ref{5.20}) we obtain
\bear
   &&I_{\rm Brane} [X^\mu,G'^{AB}] = \frac{1}{2}\int \dd^{d_c}\xi\,
    T'\, \sqrt{|G'|}\, \left ( G'^{AB} \p_A X^\mu \p_B X^\nu g_{\mu \nu}
    + 2 - d_C \right )  \nonumber \\
     && ~~~~~~~~~~~~~~~~~~~~~~~~~+ \frac{1}{d_C} \int \dd^{d_c}\xi\,
     {J^A}_\alpha {A^\alpha}_M \p_A X^M
\lbl{5.33}
\ear
The last term is just the interactive term between the conserved current
densities ${J^A}_\alpha$
and gauge potentials ${A^\alpha}_M$ coupled to
\be
   \p_A X^M \equiv \p_{a_1 ...a_r} X^{\mu_1 ...\mu_R}~,~~~~~
   r=0,1,2,...,n;~~~R=0,1,2,...,N
\lbl{5.34}
\ee
The latter potentials include the ordinary nonabelian potentials
${A^\alpha}_\mu$, $\alpha = 1,2,...,K$, coupled to $\p_A X^\mu$,
and the extra
potentials ${A^\alpha}_{\bar M}$, coupled to 
$\p_A X^{\bar M}$. In sec.\,  we argued that in a particular
parametrization of $\xi^A =(\xi,\xi^a,\xi^{a_1 a_2},...)$, if we
take $r=R=0,1,2,...,n$, we can set
\be
   \p_{a_1 ...a_r} X^{\mu_1 ... \mu_r } = 
   \p_{[a_1}X^{\mu_1}...\p_{a_r]}X^{\mu_r}
\lbl{5.34a}
\ee
If $r=1$, the above relation is automatically true, because we have
just $\p_{a_1} X^{\mu_1}$ on both sides of the equation.
If $r=0$, then we have derivative with respect to scalar parameter $\xi$,
and eq.\,(\ref{5.34a}) is automatically satisfied as well. But for
higher grades, $r=2,3,...,n$, eq.\,(\ref{5.34a}) must be imposed
as an extra condition, namely a condition that fixes a gauge.

If we assume the validity of relation (\ref{5.34a}), we find that
the interactive term in eq.\,(\ref{5.33}), namely
\bear
   I_{\rm int} &=& \frac{1}{d_C} \int \dd^{d_c}\xi\,
     {J^A}_\alpha {A^\alpha}_M \p_A X^M = \frac{1}{d_C} \int \dd^{d_c}\xi\,
     {J^A}_\alpha ({A^\alpha}_\mu
    \p_A X^\mu + {A^\alpha}_{\bar M} \p_A X^{\bar M})\nonumber \\
   &=& \frac{1}{d_C} \int \dd^{d_c}\xi\, J^{a_1 ...a_r}_\alpha
   A^\alpha_{\mu_1 ... \mu_R} \p_{a_1 ... a_r} X^{\mu_1 ...\mu_R}
\lbl{5.35}
\ear
contains the coupling of the antisymmetric gauge potentials to
the antisymmetric current density. In the case of a single
Killing ``vector" field, $\alpha=1$, the gauge fields
$A_{\mu_1 ...\mu_R}$ are abelian, and for $R=r=n$, eq.\,(\ref{5.35})
becomes the interactive term for the well known {\it Kalb-Ramond fields}
\ci{Kalb-Ramond}.
The latter fields have an important role in string theories and
brane theories\,\ci{Duff}.
Here we have demonstrated a possible broader theoretical framework
for generalized branes, coupled to generalized gauge fields,
which includes strings and Kalb-Ramond fields.
An alternative approach to generalized gauge (Maxwell) fields
in $C$-space has been considered in refs.\,\ci{CastroMaxwell}.
    
To sum up, the action (\ref{5.33}) contains the coupling of generalized
gauge fields
with the charge current density. Besides the ordinary gauge fields
${A^\alpha}_\mu$ there also occur higher grade, Kalb-Ramond
fields ${A^\alpha}_{\bar M} \equiv {A^\alpha}_{\mu_2 ... \mu_R}$
and the zero grade, scalar, field $A^\alpha$. All those fields
are included in the compact coupling term (\ref{5.35}),
where $R=0$ stands for the scalar component, $R=1$ for vector
and $R=2,3,...,N$ for higher grade components, and analogously
for $r=0,1,2,...,n$. Eq.\,(\ref{5.35}) contains
the non abelian generalization of the well
known coupling term for Kalb-Ramond fields considered in
refs.\,\ci{Kalb-Ramond,AuriliaMaxwell,Aurilia}.
The latter coupling occurs as
a special case of eq.\,(\ref{5.35}), if we take the terms with the grade
$r=R=n$ only, i.e., the terms with the maximal grade of the worldsheet
multivectors.
However, more general coupling terms with  $r \neq R \neq n$ also exists
in our theory. Thus the case for $r=n,~R=1$ includes the well
known electrically charged closed membrane considered by Dirac 
\ci{DiracMembrane}. In the following subsection we will discuss
more explicitly some particular cases of our general theory, that
were previously considered in the literature as separate subjects.

\subsection{Comparison with previous theories of charged branes}

The electromagnetic potential $A_\mu$ couples to the time like tangent
element of the worldline $X^\mu (\tau)$ of a charged {\it particle}:
\be
      I_{\mbox{\tiny int}}^{\mbox{\tiny particle}} = 
      q \int \dd \tau \, {{\p X^\mu}\oo
      {\p \tau}} \, A_\mu
\lbl{6.1}
\ee
where $q$ is the particle's electrig charge.

For extended objects the coupling involves the worldsheet tangent elements
(velocities) (\ref{2.29}) and genelarized Maxwell potentials 
$A_{\mu_1 ...\mu_n}$. The corresponding equation for a {\it string} is
\be
    I_{\mbox{\rm \tiny int}}^{\mbox{\rm \tiny string}} = 
    {{q^{12}}\oo 2!} \int \dd^2 \xi \,
    \p_{[1} X^\mu \, \p_{2]} X^\nu \, A_{\mu \nu}
\lbl{6.2}
\ee
where the two-vector charge $q^{12}$ is coupled to a two-vector potential
$A_{\mu \nu}$. For a generic {\it brane} the coupling reads
\be
    I_{\mbox{\tiny int}}^{\mbox{\tiny brane}} = {1\oo n!} q^{12...n} \int \dd^n \xi
    \, \p_{[1} X^{\mu_1}... \p_{n]} X^{\mu_n} \, A_{\mu_1 ...\mu_n}
\lbl{6.3}
\ee
Eq.(\ref{6.3}) can be rewritten as
\be    
    I_{\mbox{\tiny int}}^{\mbox{\tiny brane}} = \left ( {1\oo n!} 
    \right )^2
     q^{a_1 a_2...a_n} \int \dd^n \xi
    \, \p_{[a_1} X^{\mu_1}... \p_{a_n]} X^{\mu_n} \, A_{\mu_1 ...\mu_n}
\lbl{6.3a}
\ee    
Here $q$, $q^{a_1 a_2}$,..., $q^{a_1 a_2...a_n}$ are the coupling strengths.

If we introduce
\be
     q_n \equiv {1\oo n!} q^{a_1...a_n} \epsilon_{a_1 ... a_n}
\lbl{6.4}
\ee
and use eqs. (\ref{2.41}),(\ref{2.45}) we have
\be
    I_{\mbox{\tiny int}}^{\mbox{\tiny brane}} = 
    {1\oo n!} q_n \int \dd {\tilde \xi}
    \, {{\p X^{\mu_1 ... \mu_n}}\oo {\p {\tilde \xi}}} \, A_{\mu_1 ...\mu_n}
\lbl{6.5}
\ee

In the case considered above the vector potential $A_\mu$ couples
to the vector
tangent element of the particle's world line, the 2-vector potential
$A_{\mu \nu}$ couples to the 2-vector tangent element of the string's
world sheet, and in general, the $n$-vector potential $A_{\mu_1 ... \mu_n}$
couples to the $n$-vector tangent element of the brane's worldsheet.
By this the possible couplings are not exhausted. Long time ago Dirac
considered a relativistic charged closed membrane coupled to
the ordinary Maxwell field given in terms of the vector potential
$A_\mu$. Later, this theory has
been generalized \ci{Barut-PavsicPhase,Barut-Pavsic} to closed branes
of any dimension.

It is well known that for {\it open strings and membranes} the electric charge
$q$ ---dues to its repulsive character--- can only be concentrated on the
boundary
(e.g., at the string's ends). But for {\it closed strings} and, in general,
closed $p$-branes, $q$ can be distributed over such extended objects.
So instead
of the total charge $q$ we have to introduce the {\it charge density}.
But since the brane moves, we have the charge current density on the 
brane. In a covariant description we introduce the {\it charge current density}
$j^a$ on the brane's worldsheet $V_n$.
In ref. \ci{Barut-Pavsic} the following action was considered:
\be
    I[X^{\mu},A_\mu] = \int \dd^n \xi \, (\kappa \, 
    \sqrt{|\mbox{\rm det} \, 
    \p_a X^\mu
    \, \p_b X_\mu|} + j^a \, \p_a X^\mu \, A_\mu ) + {1\oo 4 \pi} \int
    \dd^N x \, \sqrt{|g|} \, F_{\mu \nu} F^{\mu \nu}
\lbl{6.6}
\ee
The first term in eq.(\ref{6.6}) is just the minimal surface
Dirac-Nambu-Goto 
action for a $p$-brane ($n= p+1$), whilst the second term represents
the minimal
coupling of the brane's electric charge current density $j^a$ with the
electromagnetic field potential $A_\mu$. The last term is the kinetic term
for the electromasgnetic field $F_{\mu \nu} = \p_\mu A_\nu - \p_\nu A_\mu$.

The action (\ref{6.6}) has the following transformation property:
\be
     I[X^\mu, A'_\mu] =  I[X^\mu, A_\mu] + \int \dd^n \xi \, j^a \p_a X^\mu
   \, \p_\mu \varphi =  I[X^\mu, A_\mu] + \int \dd^n \xi \, \p_a (j^a \varphi)
\lbl{6.7}
\ee
where
\be
    A'_\mu = A_\mu + \p_\mu \varphi
\lbl{6.8}
\ee
and
\be
     \p_a j^a = 0
\lbl{6.9}
\ee
Eq.(4.8) is the gauge  transformation of $A_\mu$, whilst eq.(\ref{6.9})
expresses the charge conservation. So we have
\be
     \int \dd^n \xi \, \p_a j^a = \oint \dd \Sigma_a j^a = q(\Sigma_2)
     - q(\Sigma_1) = 0
\lbl{6.9a}
\ee
where $q= \int \dd \Sigma_a \, j^a$ is the total charge of the closed brane
and $\dd \Sigma_a$ the hypersurface element.
Here $j^a$ is a worldsheet vector density (not vector), so that 
(\ref{6.9a}) is covariant under reparametrizations of $V_n$.

Here $j^a$ is the {\it intrinsic current density}, a vector density on $V_n$.
By the relation 
\be
   j^\mu = \int \dd^n \xi \, \delta (x-X(\xi)) j^a \p_a X^\mu
\lbl{6.6b}
\ee
we obtain the {\it extrinsic current
density}, i.e., the current density in the target space $V_N$. The minimal
coupling Lagrangian in (\ref{6.6}) then reads
\be
     \int \dd^n \xi \, j^a \, \p_a X^\mu \, A_\mu = 
     \int \dd^N x \, j^\mu A_\mu
\lbl{6.6a}
\ee
It is straightforward to show that the conservation law (\ref{6.9})
implies
\be
    \p_\mu j^\mu = 0
\lbl{6.6c}
\ee

The transformed action (which is a functional of $X^\mu$ and $A'_\mu$) differs
from the ``original" action $I[X^\mu, A_\mu]$ by a term with the total divergence
which has no influence on the equations of motion. The {\it canonical momentum
density} is
\be
     {p_\mu}^a = {{\cal L}\oo {\p \p_a X^\mu}} = \kappa \, \sqrt{|f|} \,
     \p^a X_\mu + j^a A_\mu
\lbl{6.10}
\ee
and it obviously is not invariant under gauge transformations (\ref{6.8}).
Therefore it is convenient to introduce the {\it kinetic momentum density}
\be
      {\pi_\mu}^a \equiv {p_\mu}^a - j^a A_\mu
\lbl{6.11}
\ee
which is gauge invariant. By employing (\ref{6.11}) we can write {\it the
phase space action} \ci{Barut-PavsicPhase}
\be
    I_{\mbox{\tiny Brane}} [X^\mu, {p_\mu}^a, \gamma_{ab}] = 
    \int \dd^n \xi \,
    \Biggl[ {p_\mu}^a \p_a X^\mu - \mbox{${1\oo 2}$} \, {{\gamma_{ab}}
    \oo {\kappa \sqrt{|\gamma|}}} \, ({\pi_\mu}^a {\pi_\nu}^b \, g^{\mu \nu}
    - \kappa^2 \, |\gamma| \gamma^{ab}) + \kappa \sqrt{|\gamma|} \, 
    (1-n) 
    \Biggr]
\lbl{6.12}
\ee
After eliminating ${p_\mu}^a$ from its equations of motion
\be
    {p^\nu}_c - j_c A^\nu = \kappa \sqrt{|\gamma|} \, \p_c X^\nu
\lbl{6.13}
\ee
we obtain \ci{Barut-Pavsic,PavsicBook} the Howe-Tucker action in the presence
of
electromagnetic field:
\be
    I_{\mbox{\tiny Brane}} [X^\mu, \gamma_{ab}] = 
    \int \dd^n \xi \Biggl[ 
    {{\kappa \, \sqrt{|\gamma|}}\oo 2} (\gamma^{ab} \p_aX^\mu \, \p_b X_\mu
    +2-n) + j^a \p_a X^\mu \, A_\mu \Biggr]
\lbl{6.14}
\ee
The total action action is thus the sum of the brane action 
$I_{\mbox{\tiny Brane}}$
and the kinetic term for the electromagnetic field 
$I_{\mbox{\tiny EM}}$ (given by the last term in (\ref{6.6}).

A closed brane can thus possess two different kinds of charges coupled to
two different kinds of gauge fields:
\begin{description}
     \item[(i)] Kalb-Ramond charge $q^{12...n}$ coupled to Kalb-Ramond gauge
     field potentials $A_{\mu_1 ...\mu_n}$ according to eq.(\ref{6.3});
     
     \item[(ii)] the ordinary electric charge $q$ coupled to the ordinary
     Maxwell potential $A_\mu$ according to eq.(\ref{6.6}) (or equivalently,
     eqs. (\ref{6.12}) and (\ref{6.14})).
\end{description}

By inspecting the interactive action (\ref{6.3}) we observe that
it is just a
coupling term that can be aded to the action (\ref{2.35a}). On the other
hand, the action (\ref{6.14}) is just a generalization of the action
(\ref{2.31}) to which we added the minimal coupling term for the
Maxwell field
$A_\mu$. Both actions, (\ref{2.31}) and (\ref{2.35a}), were generalized
to $C$-space, and so we have obtained the action (\ref{2.67}) (which is
equivalent to the action (\ref{2.80})). As we have seen in Sec.\,5.3,
the two kinds of couplings, namely
(i) and (ii), can be unified by employing the $C$-space description,
in which they arise from the metric of the target $C$-space $C_{V_N}$,
equipped with connection, whose curvature was assumed to
be in general different from zero.       

\section{Discussion}

Spacetime as a continuum of points is just a start, from
which we can arrive at a continuum of oriented areas, also called oriented
volumes. Such enlarged continuum, called Clifford space or $C$-space
provides a framework for generalizing the concepts of event, point particle,
string and, in general, brane. As an ordinary brane is an object
that extends in spacetime,
so a generalized brane is an object that extends in $C$-space. An
important feature of $C$-space is that oriented areas (volumes) of different
grades, associated with branes of different
dimensionalities, can be transformed into each other. A generic
geometric object has mixed grade, it is represented by a Clifford number,
also called Clifford aggregate or polyvector, and it is associated
with a generalized brane. Having set such a kinematics, one can
construct a dynamics which employs the well known Howe-Tucker brane
action, which is now generalized to $C$-space. The latter space can
be curved, and this gives rise, \` a la Kaluza-Klein, to gauge fields
as components of the $C$-space metric. The $C$-space Howe-Tucker action,
minimally coupled to the $C$-space metric, contains, in particular,
the well known terms for the coupling of a $p$-brane (including a point particle)
to gauge fields. Amongst the latter gauge fields there also occur
the Kalb-Ramond antisymmetric fields, and their non Abelian
generalizations\,\ci{Savidy}.
In this paper we focused our attention to the dynamics of the
branes, and left aside the fact that in a more complete
treatment\,\ci{PavsicKaluza,PavsicKaluzaLong} $C$-space itself becomes dynamical,
as in general relativity.

In the present paper we presented only a piece of the story, namely, how
the classical theory of branes in a fixed background could be generalized
to Clifford space. Since with the points of a flat Clifford space one
can associate Clifford numbers (polyvectors), that are elements of  Clifford
algebra ${\bfs C}_N$, this automatically brings spinors
(as members of left or right ideals of ${\bfs C}_N$)
into our description. A polyvector $X^{\mu_1... \mu_R} \gam_{\mu_1 ... \mu_R}$,
since it can be rewritten, e.g., in terms of a basis spanning all
independent left ideals,
thus contains spinorial degrees of freedom 
\ci{PavsicKaluza,PavsicKaluzaLong,PavsicSaasFee}.
This means that by describing our branes in terms of the $C$-space embedding
functions $X^M \equiv X^{\mu_1... \mu_R}$ we have already included spinorial
degrees of freedom. We do not need to postulate them separately, as is
done in ordinary string and brane theories, where besides Grassmann even
( ``bosonic") variables $X^\mu$, there occur also Grassmann odd
(``fermionic") or spinorial variables. In this formulation
we have a possible clue to the resolution of a big open problem,
namely, what exactly is string theory. I believe that further
research into that direction would provide very fruitful results
and insight. An important insight is already in recognizing that
16-dimensional Clifford space provides a consistent
description of quantized string theory\,\ci{PavsicParis,PavsicSaasFee}.
The underlying spacetime can remain 4-dimensional, there is no need
for a 26-dimensional or a 10-dimensional spacetime. The extra degrees of
freedom required for consistency of string theory, described in terms
of variables $X^M (\tau,\sigma)$, are due to
extra dimensions of $C$-space, and they need not be compactified; they are
due to volume (area) evolution, and are thus all physical.
But since a generic component $X^M (\tau,\sigma) 
\equiv  X^{\mu_1... \mu_R}(\tau,\sigma)$ denotes an
oriented $R$-volume, associated with an $(R-1)$-brane
(i.e., a $p$-brane for $p=R-1$), we have that string itself (i.e., 1-brane)
is not enough for consistency. Higher branes are automatically present
in the description with functions $X^{\mu_1 ...\mu_R} (\tau,\sigma)$,
although they are not described in full detail, but only up to the
knowledge of oriented $R$-volume. Because of the presence of two
parameters $\tau,~\sigma$, we keep on talking about evolving {\it strings},
not in 26 or 10-dimensional spacetime, but in 16-dimensional
Clifford space. In general, the number of parameters can be arbitrary,
but less then 16, and so we have a brane in $C$-space, i.e.,
a generalized brane discussed in this paper.

The $C$-space approach to branes is possibly related to
the approach with extra coordinates, considered by Duff and Lu\,\ci{DuffDual}
in order to describe brane dualities. 
Such interesting relation needs to be investigated in future research.

Clifford algebras in infinite dimensions and in continuous dimensions
are still an uncharted territory worth exploring. A pioneering step
into that direction has been done in ref.\,\ci{PavsicBook}, where a
theory of generalized branes was formulated in terms of the generators of the
infinite dimensional Clifford algebra.

Another possible direction that remains to be further explored is
related to the fact that p-brane actions can be recast as non Abelian
gauge theories of volume preserving diffeomorphisms\,\ci{Aurilia, CastroAbel}.
It would be interesting to extend such approach to the case of the
generalized branes in $C$-space considered in this paper.
 Also there is a number of works on polyvector generalized supersymmetries
\,\ci{CastroSuper}, on the implications to M theory\,\ci{CastroMtheory},
on generalized Yang-Mills theories, Poly-particles and duality in $C$-spaces
\,\ci{CastroYang,Savidy}
and on quantum mechanics in $C$-spaces and non-commutativity\,\ci{CastroQuant}
A possible connection of $C$-space to twistors (see, e.g.,
\ci{Penrose}) would also be worth investigating.

Clifford algebra, without recourse to $C$-space and generalized branes,
has been considered in numerous works attempting to explain the standard
model \ci{StandardCliff}. There are diverse approaches, but common to all
of them is a feeling that Clifford algebra might be a clue to the unification
of fundamental forces. It is reasonable to expect that further research
will bring useful results, and that it will be crucial to take into account
the concepts of $C$-space and branes .

\vs{5mm}  

\centerline{Acknowledgement}

This work was supported by the Ministry of
High Education, Science and Technology of Slovenia.

{\small

\end{document}